\documentclass[]{aastex631}

\usepackage{graphicx}   
\usepackage{amsmath,amssymb}
\usepackage{bm}
\usepackage{physics}
\usepackage{comment}

\graphicspath{{./}{figures/}}

\accepted{10 October 2025, Journal of Computational Physics}

\shorttitle{Special Relativistic Godunov SPH}
\shortauthors{Kitajima, Inutsuka \& Seno}
\begin{document}

\title{Special Relativistic Smoothed Particle Hydrodynamics Based on Riemann Solver}

\author[0000-0003-0738-0707]{Kanta Kitajima}
\altaffiliation{Email: {\tt kitajima.kanta.p4@a.mail.nagoya-u.ac.jp}}
\affiliation{Department of Physics, \\
Graduate School of Science, \\
Nagoya University, \\
Furo-cho, Chikusa-ku, Nagoya  464-8692, Japan}

\author[0000-0003-4366-6518]{Shu-ichiro Inutsuka}
\affiliation{Department of Physics, \\
Graduate School of Science, \\
Nagoya University,  \\
Furo-cho, Chikusa-ku, Nagoya  464-8692, Japan}

\author[0009-0003-9296-3161]{Izumi Seno}
\affiliation{Department of Physics, \\
Graduate School of Science, \\
Nagoya University,  \\
Furo-cho, Chikusa-ku, Nagoya  464-8692, Japan}

\begin{abstract}
This paper proposes a novel numerical method based on Godunov Smoothed Particle Hydrodynamics for special relativistic fluid dynamics. Our method utilizes a Riemann solver to describe shock, enhancing accuracy in strong shock waves.
The formulation maintains conservation laws and achieves higher accuracy through convolution integrals that define physical quantities for SPH particles.
We also propose the number density calculation method that uses a non-equal baryon number in each SPH particle and variable smoothing length in a way different from the conventional method. Numerical experiments demonstrate the method’s robustness across one- and two-dimensional relativistic shock tube problems, as well as its ability to simulate Kelvin-Helmholtz instabilities accurately, validating SRGSPH as a reliable approach for high-resolution relativistic simulations.
\end{abstract}
\section{Introduction}
\label{sec:intro}
Smoothed Particle Hydrodynamics (SPH) is a widely used computational method in fluid dynamics, distinguished by its Lagrangian framework that represents fluid as a collection of discrete pieces, referred to as SPH particles.  \citep[e.g.,][]{Gingold_Monaghan_1977,Lucy_1977}.
The notable adaptability of SPH in simulating diverse physical phenomena, including self-gravity, radiative cooling, and chemical reactions, has contributed to its extensive deployment in the domain of astrophysics. 
One of the primary advantages of SPH is its capacity to accommodate significant deformations without the necessity of a computational grid. 
These capabilities render SPH a convenient tool for investigating various astrophysical phenomena, such as the formation of galaxies, stars, and planets \citep[e.g.,][]{Springnel_2010,Monaghan_2012}.
The extension to magnetohydrodynamics has also been well developed by many authors\citep[e.g.,][]{Iwasaki_Inutsuka_2011,Price_2012,Tsukamoto_etal_2013}.
Especially, \citet{Tsukamoto_etal_2017,Tsukamoto_etal_2021} have included all the three non-ideal magnetohydrodynamics effects successfully. 
The extension to elastic dynamics is also developed\citep{Sugiura_Inutsuka_2017}. 
In particular, relativistic SPH is crucial for simulating extreme cosmic events, such as tidal disruption events and collisions between compact objects \citep[e.g.,][]{Rosswog+_2008,Rosswog_2013}.

Special Relativistic SPH (SRSPH) builds upon the standard SPH method by incorporating the principles of special relativity to model high-energy and high-velocity astrophysical events.
A useful review of SRSPH can be found in \citep{Rosswog_2015} (see also \citep[e.g.,][]{Chow_Monaghan_1997,Rosswog_2010}.
Conventional SRSPH techniques, which frequently employ a constant artificial viscosity to calculate shock waves, encounter difficulties when confronted with multiple shock waves of varying intensities. A single, fixed artificial viscosity parameter can result in excessive smoothing of weaker shocks, thereby reducing the precision of the simulation. To overcome these limitations, adaptive methodologies that dynamically adjust the artificial viscosity in accordance with shock intensity have been proposed. However, these approaches often lead to increased complexity in the simulation.

In this context, we introduce a novel approach: Special Relativistic Smoothed Particle Hydrodynamics based on Godunov’s method (SRGSPH). This method, an extension of the GSPH method \citep{Inutsuka_2002}, employs a Riemann solver, an approach initially developed for grid-based methods, to enhance the accuracy in shock wave simulations. The Riemann solver enables SRGSPH to handle shock waves with variable intensity more effectively, allowing for precise control of artificial viscosity and improved handling of strong shocks. Furthermore, SRGSPH incorporates convolution integrals for computing the physical quantities of SPH particles, thereby ensuring a smoother and more accurate representation of fluid properties in a relativistic framework.

Another critical innovation in SRGSPH is its approach to calculating particle number density, particularly in cases with non-uniform particle baryon numbers and variable smoothing lengths. Unlike conventional methods, SRGSPH employs a volume-based approach, allowing for a more accurate representation of spatially varying densities and preventing common artificial problems seen in density discontinuities with non-uniform particle baryon numbers.
Based on this method, the introduction of variable smoothing lengths also enhances SRGSPH’s ability to handle low-density regions, a capability essential for accurately modeling astrophysical environments where particle spacing can vary significantly.
This approach can also apply to non-relativistic SPH.

In the next section, we derive the formulation of SRGSPH.
Results of a series of tests are presented in \S3.
\S4 presents a summary.

\section{Method}
\label{sec:methods}

\subsection{Basic Equations of Special Relativistic Hydrodynamics}
Bold font represents a three-dimensional vector in space.
The fundamental equations of a special relativistic perfect fluid in the Lagrangian description are given by:
\begin{align}
\label{eq:EoC}
&\frac{dN}{dt}=-N\nabla\cdot\bm{v},\\
\label{eq:EoM}
&\frac{d\bm{S}}{dt}=-\frac{1}{N}\nabla P,\\
\label{eq:EoE}
&\frac{de}{dt}=-\frac{1}{N}\nabla\cdot (P\bm{v}),
\end{align}
where $N$, $\bm{S}$, and $e$ are the baryon number density, the relativistic canonical momentum per baryon, and the canonical energy per baryon in the laboratory frame.
Lagrangian derivative is defined as follows:
\begin{align}
\frac{d}{dt} \equiv \partial_t + \bm{v}\cdot\nabla.
\end{align}
Furthermore, $\bm{S}$ and $e$ are given as follows:
\begin{align}
\label{eq:S}
  \bm{S}   &= \gamma H \bm{v},\\
\label{eq:e}
  e &= \gamma H -\frac{P}{Nc^2},
\end{align}
where $\gamma$ and $H$ are the Lorentz factor of the fluid element and the enthalpy per baryon, and they are given as follows:
\begin{align}
  \gamma &= \sqrt{\frac{1}{1-\bm{v}^2/c^2}},\\
  H      &= 1 + \frac{u}{c^2} + \frac{P}{nc^2},
\end{align}
where $u$ and $n$ are the thermal energy per baryon and the baryon number density in the rest frame, and $N=\gamma n$ holds.

To obtain a solution for fluid motion, it is necessary to have an equation of state in addition to the equations of continuity, motion, and energy.
In this paper, we use the equation of state of an ideal gas, which is as follows:
\begin{align}
\label{eq:EoS}
    P = (\gamma_c-1)nu,
\end{align}
where $\gamma_c$ is the ratio of specific heats.
Other symbols have their usual meanings.

\subsection{Formulation of SRGSPH with Fixed Smoothing Length}
\subsubsection{Basic Definitions}
In this section, we assume that the smoothing length is constant and does not vary over time for simplicity.
The case where it changes over time will be discussed in the next section.
Furthermore, we assume that the number of particles constituting the fluid within each SPH particle remains constant.

In the SPH method, it is necessary to define a field in order to calculate gradient quantities such as the pressure gradient force.
Regardless of whether the context is relativistic or non-relativistic, many SPH methods place special emphasis on the number density (or simply density), similar to kernel density estimation, and define it as follows:
\begin{align}
    N(\bm{x})\equiv \nu \sum_j W(\bm{x}-\bm{x}_j,h),
\end{align}
where $\bm{x}_j$ is the position of an SPH particle, $\nu$ is the baryon number in an SPH particle and $W$ is a kernel function and we can use various kernel functions in our formulation but we use the following Gaussian kernel in this article just for simplicity:
\begin{align}
    W(\bm{x},h) = \left[ \frac{1}{h \sqrt{\pi}} \right]^d\exp \left[ -\frac{\bm{x}^2}{h^2} \right],
\end{align}
where $d$ and $h$ are the number of dimensions and the smoothing length.
In this section, we consider a fixed smoothing length, which is generally preferred to be approximately the average particle spacing.

Next, we define the physical quantities at the position of each SPH particle.
We define the number density and the other physical quantities that the $i$-th SPH particle has by the convolution of the kernel function as follows:
\begin{align}
    \ev{N_i}&\equiv N(\bm{x}_i),\\
    \label{eq:convolution_consth}
    \ev{f_i}&\equiv \int f(\bm{x})W(\bm{x}-\bm{x}_i,h)d\bm{x}.
\end{align}
We estimate the error between the field quantities and those carried by the SPH particles.
Substituting the Taylor expansion of $f(\bm{x})$ around $\bm{x} = \bm{x}_i$ into Eq.(\ref{eq:convolution_consth}), we obtain
\begin{align}
    \label{eq:accuracy}
    \ev{f_i} = f(\bm{x}_i) + \frac{h^2}{4}f''(\bm{x}_i) + \mathcal{O}(h^4).
\end{align}
The difference of $f(x_i)$ and $\ev{f_i}$ is on the order of $h^2$.

\subsubsection{Basic equations}
In the basic equations of the SRGSPH method, we use the equation of the number density of SPH particles instead of the equation of continuity and derive the equations of motion and energy by convolution with the kernel. In this subsection, we derive the equations of motion and energy. In the SPH method, the physical quantities of actual fluid and the values that the SPH particle has are considered equal, i.e., we do not care about the presence or absence of brackets in the actual calculations.
We obtain the equation of motion through the convolution of Eq.(\ref{eq:EoM})'s two sides with the kernel function as follows:
\begin{align}
  \ev{\dot{\bm{S}}_i}&\equiv  \int  \frac{d\bm{S}(\bm{x})}{dt}W(\bm{x}-\bm{x}_i,h)d\bm{x}\\
  &=-\int \frac{1}{N(\bm{x})}\nabla P(\bm{x}) W(\bm{x}-\bm{x}_i,h)d\bm{x}\\
  &=-\int\left[ \nabla\frac{P(\bm{x})}{N(\bm{x})} + \frac{P(\bm{x})}{N^2(\bm{x})}\nabla N(\bm{x}) \right] W(\bm{x}-\bm{x}_i,h)d\bm{x}\\
  &=+\sum_j\int \frac{\nu P(\bm{x})}{N^2(\bm{x})} \left[W(\bm{x}-\bm{x}_j,h) \nabla W(\bm{x}-\bm{x}_i,h) - \nabla W(\bm{x}-\bm{x}_j,h) W(\bm{x}-\bm{x}_i,h) \right] d\bm{x}\\
  &=-\sum_j\int \frac{\nu P(\bm{x})}{N^2(\bm{x})} [\nabla_i - \nabla_j] W(\bm{x}-\bm{x}_i,h) W(\bm{x}-\bm{x}_j,h) d\bm{x}.
\end{align}
In the course of the derivation, we have utilized the property of the Gaussian function, which allows us to set the surface term’s integral over the entire space to zero.
Additionally, the following relation is used:
\begin{align}
    \nabla W(\bm{x}-\bm{x}_j,h) = -\nabla_j W(\bm{x}-\bm{x}_j,h).
\end{align}

The energy equation can be derived in a similar manner to the equation of motion, as follows:
\begin{align}
  \ev{\dot{e}_i}
  &\equiv  \int \frac{de(\bm{x})}{dt}W(\bm{x}-\bm{x}_i,h)d\bm{x}\\
  &=-\int \frac{1}{N(\bm{x})}\nabla \cdot [P(\bm{x})\bm{v}(\bm{x})] W(\bm{x}-\bm{x}_i,h)d\bm{x}\\
  &=-\int \left[ \nabla \cdot \frac{P(\bm{x})\bm{v}(\bm{x})}{N(\bm{x})} + \frac{P(\bm{x})\bm{v}(\bm{x})}{N^2(\bm{x})}\cdot \nabla N(\bm{x}) \right]W(\bm{x}-\bm{x}_i,h)d\bm{x}\\  
  &=+ \sum_j \int \frac{\nu P(\bm{x})\bm{v}(\bm{x})}{N^2(\bm{x})} \cdot [W(\bm{x}-\bm{x}_j,h) \nabla W(\bm{x}-\bm{x}_i,h) - \nabla W(\bm{x}-\bm{x}_j,h) W(\bm{x}-\bm{x}_i,h)] d\bm{x}\\   
  &=- \sum_j \int \frac{\nu P(\bm{x})\bm{v}(\bm{x})}{N^2(\bm{x})} \cdot [\nabla_i - \nabla_j] W(\bm{x}-\bm{x}_i,h) W(\bm{x}-\bm{x}_j,h) d\bm{x}.
\end{align}

\subsubsection{Evaluation of convolution integrals}\label{evaluation_const}
Both the equations of motion and energy contain a similar integral term.
This term depends on the position of SPH particles, so can't be calculated analytically.
Therefore, the integral appearing in the above equation is transformed as follows:
\begin{align}
    &\int \frac{f(\bm{x})}{N^2(\bm{x})}[\nabla_i-\nabla_j] W(\bm{x}-\bm{x}_i,h)W(\bm{x}-\bm{x}_j,h)d\bm{x}\\
    &=\int \frac{f(\bm{x})}{N^2(\bm{x})}[\nabla_i-\nabla_j]\left( \frac{1}{h\sqrt{\pi}} \right)^{2d}\exp\left[ -\frac{ 2\left( \bm{x}-\frac{\bm{x}_i+\bm{x}_j}{2} \right)^2+\frac{(\bm{x}_i-\bm{x}_j)^2}{2} }{h^2} \right]d\bm{x}\\
    &= f_{ij}V^2_{ij}[\nabla_iW(\bm{x}_i-\bm{x}_j,\sqrt{2}h) - \nabla_jW(\bm{x}_i-\bm{x}_j,\sqrt{2}h)],
\end{align}
where
\begin{align}
    V^2_{ij}(h)&\equiv\int \frac{1}{N^2(\bm{x})} \left( \frac{\sqrt{2}}{h\sqrt{\pi}} \right)^d \exp\left[ -\frac{2}{h^2}\left( \bm{x}-\frac{\bm{x}_i+\bm{x}_j}{2} \right)^2 \right] d\bm{x},\\
    f_{ij}(h)&\equiv\frac{1}{V^2_{ij}}\int \frac{f(\bm{x})}{N^2(\bm{x})} \left( \frac{\sqrt{2}}{h\sqrt{\pi}} \right)^d \exp\left[ -\frac{2}{h^2}\left( \bm{x}-\frac{\bm{x}_i+\bm{x}_j}{2} \right)^2 \right] d\bm{x}.
\end{align}
Since $N(\bm{x})$ is obtained from the distribution of SPH particles, $V^2_{ij}(h)$ cannot be solved analytically.
It is possible to perform the integration numerically, but the computational cost would be enormous.
By interpolating $N^{-2}(\bm{x})$ in $V^2_{ij}(h)$ such as linear or cubic spline (with careful attention to preserving the symmetry between  $i$  and  $j$ , as it is crucial for conservation laws), the integral can be solved analytically.
Here, $V_{ij, {\rm interp}}^2$ corresponds to the approximate $N^{-2} (\bm{x})$ that appears as a result of the interpolation and integration.
This method is described in \citep{Inutsuka_2002}, so we omit this in this paper.

Also, to accurately describe shock waves, we use 
the result of solving the Riemann problem between particles $i$ and $j$, 
$f^*_{ij}$ instead of $f_{ij}(h)$.
As a result, SRGSPH's equations of motion and energy are expressed as follows;
\begin{align}
\label{eq:SRGSPH_EoM_const}
    \ev{\dot{\bm{S}}_i} &= -\sum_j \nu P^*_{ij}V^2_{ij,\, {\rm interp}}[\nabla_iW(\bm{x}_i-\bm{x}_j,\sqrt{2}h) - \nabla_jW(\bm{x}_i-\bm{x}_j,\sqrt{2}h)], \\
\label{eq:SRGSPH_EoE_const}
    \ev{\dot{e}_i} &= -\sum_j \nu P^*_{ij}\bm{v}^*_{ij}\cdot V^2_{ij,\, {\rm interp}}[\nabla_iW(\bm{x}_i-\bm{x}_j,\sqrt{2}h) - \nabla_jW(\bm{x}_i-\bm{x}_j,\sqrt{2}h)].
\end{align}
Eqs.(\ref{eq:SRGSPH_EoM_const}) and (\ref{eq:SRGSPH_EoE_const}) satisfy the anti-symmetry of the indices $i$ and $j$ because $P^*_{ij}=P^*_{ji}$, $\bm{v}^*_{ij}=\bm{v}^*_{ji}$, $V^2_{ij}=V^2_{ji}$ and $\nabla_iW(\bm{x}_i-\bm{x}_j,\sqrt{2}h)=-\nabla_jW(\bm{x}_j-\bm{x}_i,\sqrt{2}h)$ hold, and show that total momentum and energy are conserved.

\subsection{Formulation of SRGSPH with Variable Smoothing Length}
Next, we consider the case where the smoothing length varies over time and the baryon number carried by each SPH particle also varies depending on the particle.
In the case of a fixed smoothing length, as the particle spacing increases, particles may cease to interact with each other, making it impossible to compute low-density regions. To prevent this, it is common to use a variable smoothing length that adjusts according to the particle spacing.
Instead of focusing on number density or mass density, we emphasize the volume and define it as follows:
\begin{align}
    \label{eq:def_Vp}
    V_{\rm p}(\bm{x})\equiv\left[ \sum_j W(\bm{x}-\bm{x}_j,h(\bm{x}))\right]^{-1},
\end{align}
where, similar to the case with a fixed smoothing length, $W$ is the Gaussian kernel:
\begin{align}
    W(\bm{x},h(\bm{x})) = \left[ \frac{1}{h(\bm{x}) \sqrt{\pi}} \right]^d\exp \left[ -\frac{\bm{x}^2}{h^2(\bm{x})} \right].
\end{align}
The smoothing length $h(\bm{x})$ is defined as follows:
\begin{align}
    \label{eq:def_h}
    h(\bm{x}) &\equiv  \eta {V_{\rm p}^*}^{1/d}(\bm{x}),\\
    \label{eq:def_Vp*}
    V_{\rm p}^*(\bm{x}) &\equiv \left[ \sum_j W(\bm{x}-\bm{x}_j,C_{\rm smooth}h(\bm{x}))\right]^{-1}.
\end{align}
Note that the smoothing length $h(\bm{x})$ is defined over the entire spatial domain.
As we will see later in Section \ref{evaluation}, we implement $\eta$ and $C_{\rm smooth}$ to reduce the gradient of $h(\bm{x})$ for calculation. In this paper, we use $\eta=1.0$ and $C_{\rm smooth}=2.0$.

As you can see from Eqs.(\ref{eq:def_h}) and (\ref{eq:def_Vp*}), $h(\bm{x})$ and $V^*_{\rm p}(\bm{x})$ have an iterative relationship.
In the actual calculation, we derive the values of $V_{\rm p}(\bm{x})$ and $h(\bm{x})$ by iterating Eqs.(\ref{eq:def_h}) and (\ref{eq:def_Vp*}) until convergence only in the first step.  For the subsequent time steps, we use the previous step value of $h(\bm{x})$ to find the value of $V^*_{\rm p}(\bm{x})$.
The number density is obtained as follows:
\begin{align}
    \label{eq:N(x)}
    N(\bm{x})=\nu(\bm{x}) V_{\rm p}^{-1}(\bm{x}),
\end{align}
where $\nu(x)$ is the baryon number.

Next, we define the physical quantities for each SPH particle.
We define the number density and the other physical quantities that the $i$-th SPH particle has by the convolution of the kernel function as follows:
\begin{align}
    \ev{N_i}&\equiv N(\bm{x}_i)\\
    \label{eq:convolution}
    \ev{\nu_if_i}&\equiv \int \nu(\bm{x})f(\bm{x})W(\bm{x}-\bm{x}_i,h(\bm{x}))d\bm{x}.
\end{align}
In the case of a variable smoothing length, it is not possible to conduct a rigorous discussion as shown in Eq. (\ref{eq:accuracy}).

\subsection{Why Particle Volume is Defined?}
\begin{figure}[htbp]
  \begin{center}
   \includegraphics[width=0.95\linewidth]{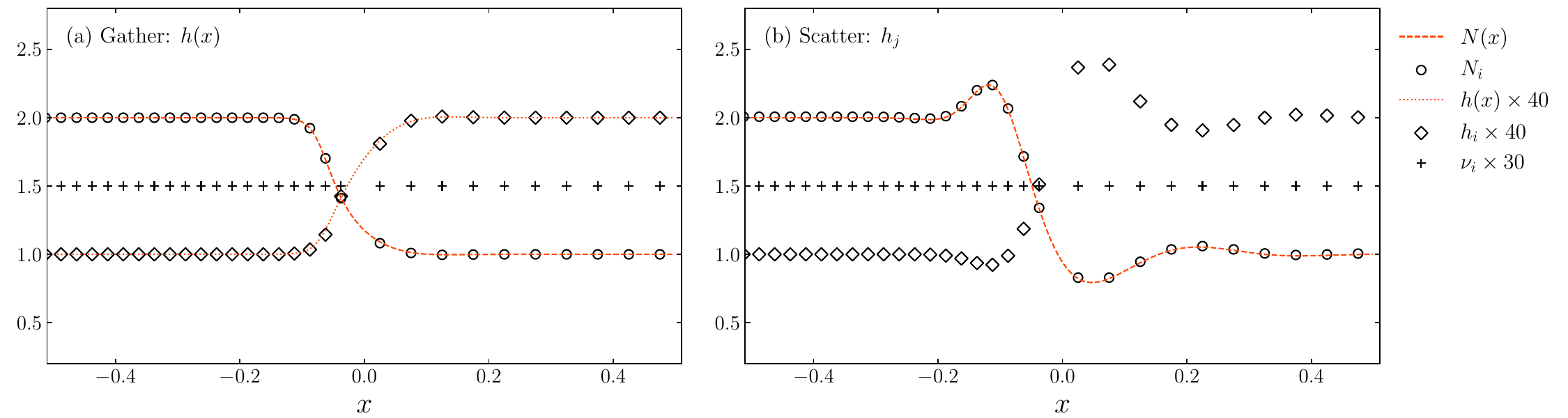}
  \end{center}
   \caption{
    The spatial profiles of the baryon number, smoothing length, and number density profile in the lab frame for each SPH particle in (a) gather and (b) scatter approaches.
    As field quantities, the density in the lab frame is represented by the dashed line and the smoothing length by the dotted line.
    The gather method defines the density in the lab frame and smoothing length as field quantities, whereas the scatter method defines only the density in the lab frame as a field quantity and does not specify the smoothing length as a field quantity.
   \label{fig:gather_vs_scatter}
   }
\end{figure}

In this section, we will discuss the definition of number density and the definition of the smoothing length in cases where a baryon number in an SPH particle and smoothing length vary.
First, we consider the case where a baryon number in an SPH particle is constant.
When using a variable smoothing length, the definition of the spatial distribution of number density can be considered in two ways: the ‘gather’ approach, which defines it using the particle’s own smoothing length, and the ‘scatter’ approach, which defines it using the smoothing length of the neighboring particles as follows:
\begin{align}
    & N_{\text {gather }}(\bm{x})\equiv \nu \sum_j W\left(\bm{x}-\bm{x}_j, h(\bm{x})\right), \\
    & N_{\text {scatter }}(\bm{x})\equiv \nu \sum_j W\left(\bm{x}-\bm{x}_j, h_j\right).
\end{align}
The smoothing length must be smoothed because the gather method introduces a $\nabla h(x)$ term into the equations of motion and energy. 
On the other hand, the scatter approach does not produce a $\nabla h(x)$ term, but it tends to cause overshooting or undershooting at discontinuities (Fig. \ref{fig:gather_vs_scatter}).
In this study, the calculation of the number density carried by SPH particles is performed using the gather approach.
To simplify the equations of motion and energy, we propose a method to ensure that the smoothing length changes smoothly, thereby allowing us to ignore the $\nabla h(x)$ term, as will be discussed later.

\begin{figure}[htbp]
  \begin{center}
   \includegraphics[width=0.95\linewidth]{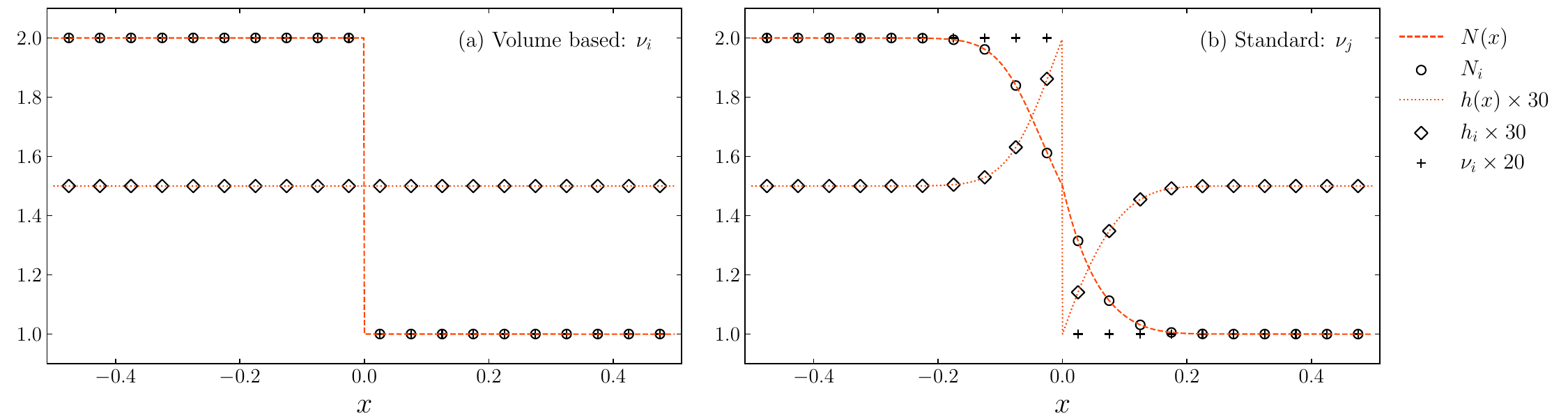}
  \end{center}
   \caption{
    The spatial profiles of the baryon number, smoothing length, and number density profile in the lab frame for each SPH particle under the volume-based and the standard approaches. 
   \label{fig:Vb_vs_Stn}
   }
\end{figure}

Next, we consider the case where the baryon number for each SPH particle varies.
In this case, there are two approaches for calculating the number density: one uses the particle’s own baryon number, and the other uses the baryon number as a field quantity.
If we refer to the former as ‘volume-based’ and the latter as ‘standard’, they can be expressed as follows:
\begin{align}
    & N_{\text {volume-based }}(\bm{x})\equiv \nu(\bm{x}) \sum_j W\left(\bm{x}-\bm{x}_j, h(\bm{x})\right),\label{eq:vb} \\
    & N_{\text {standard }}(\bm{x})\equiv \sum_j \nu_j W\left(\bm{x}-\bm{x}_j, h(\bm{x})\right).\label{eq:stn}
\end{align}
Based on this, we consider the value of density in the lab frame and smoothing length for cases with different baryon numbers. 
Fig. \ref{fig:Vb_vs_Stn} shows that the number density in the lab frame and smoothing lengths on the basis of Eqs. (\ref{eq:def_h}), (\ref{eq:def_Vp*}), (\ref{eq:vb}) and (\ref{eq:stn}).
As is evident from the figure, with the volume-based approach, the smoothing length is a constant value of particle spacing degree, while with the standard approach,  $h$  becomes sharper.
This indicates that the influence of the  $\nabla h$  term increases, leading to larger errors. 
Furthermore, as will be discussed below, a smaller $h$ leads to a smaller time step $\Delta t$, which makes longer calculations more difficult.
Therefore, in this study, we adopt the volume-based approach.

\begin{figure}[htbp]
  \begin{center}
   \includegraphics[width=0.95\linewidth]{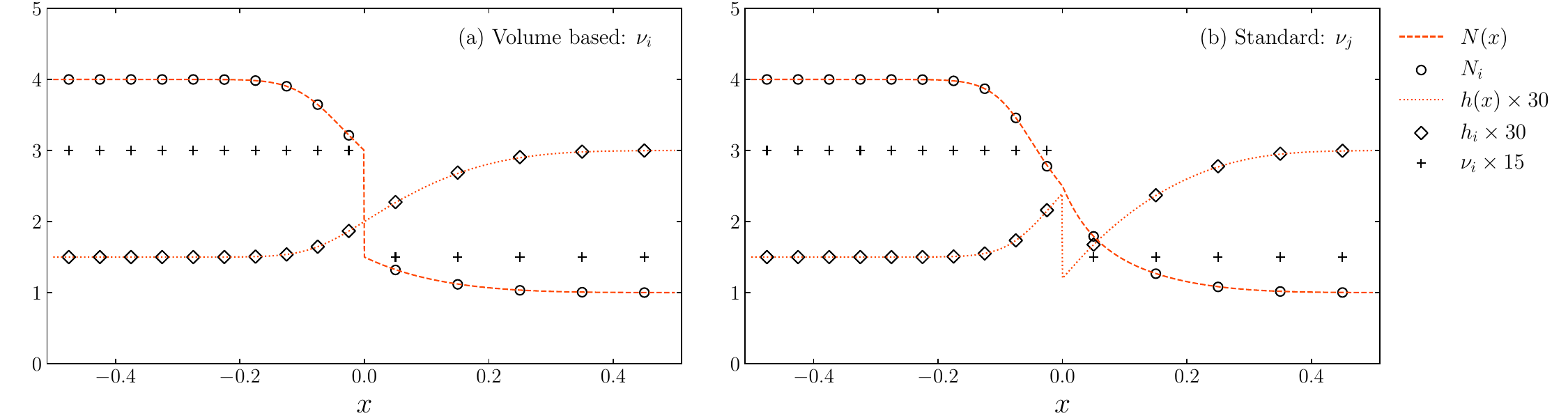}
  \end{center}
   \caption{
    The spatial profiles of the baryon number, smoothing length, and number density in the lab frame for each SPH particle in a general case where both the baryon number and smoothing length vary spatially.
In the volume-based approach, the smoothing length varies smoothly across the interface, in contrast to the standard approach, where it exhibits a discontinuous jump.
   \label{fig:Vb_vs_Stn_2}
   }
\end{figure}

In general, not only the baryon number but also the particle separation spatially varies.
Fig. \ref{fig:Vb_vs_Stn_2} shows an example that corresponds to this situation.
The smoothing length in the volume-based approach preserves spatial monotonicity, whereas in the standard approach it does not.

\subsection{Basic equations}
As with fixed smoothing lengths,
we obtain the equation of motion through the convolution of both sides of Eq.(\ref{eq:EoM}) with the kernel function.
\begin{align}
  &\ev{\nu_i\dot{\bm{S}}_i}\\
  &\equiv  \int \nu(\bm{x}) \frac{d\bm{S}(\bm{x})}{dt}W(\bm{x}-\bm{x}_i,h(\bm{x}))d\bm{x}\\
  &=-\int V_{\rm p}(\bm{x})\nabla P(\bm{x}) W(\bm{x}-\bm{x}_i,h(\bm{x}))d\bm{x}\\
  &=+\int P(\bm{x})\nabla [V_{\rm p}(\bm{x}) W(\bm{x}-\bm{x}_i,h(\bm{x}))] d\bm{x}\\
  &=+\sum_j\int P(\bm{x})V_{\rm p}^2(\bm{x})[W(\bm{x}-\bm{x}_j,h(\bm{x}))\nabla W(\bm{x}-\bm{x}_i,h(\bm{x}))-W(\bm{x}-\bm{x}_i,h(\bm{x}))\nabla W(\bm{x}-\bm{x}_j,h(\bm{x}))] d\bm{x}\\
  &=-\sum_j\int P(\bm{x})V_{\rm p}^2(\bm{x}) \left[ (\nabla_i-\nabla_j)-\frac{h}{2}\nabla h (\nabla_i^2 -\nabla_j^2) \right] W(\bm{x}-\bm{x}_i,h(\bm{x}))W(\bm{x}-\bm{x}_j,h(\bm{x}))d\bm{x}.
\end{align}
In the above manipulation, the following relation is used:
\begin{align}
    \nabla W(\bm{x}-\bm{x}_i,h(\bm{x})) = -\left[ \nabla_i - \nabla h\left( \frac{h}{2} \nabla_i^2 + \frac{1-d}{h}\right)\right] W(\bm{x}-\bm{x}_i,h(\bm{x})).
\end{align}
The anti-symmetry of the indices $i$ and $j$ in this equation guarantees the action-reaction principle.

The energy equation is derived in a similar way.
\begin{align}
  &\ev{\nu_i\dot{e}_i}\\
  &\equiv  \int \nu(\bm{x}) \frac{de(\bm{x})}{dt}W(\bm{x}-\bm{x}_i,h(\bm{x}))d\bm{x}\\
  &=-\int V_{\rm p}(\bm{x})\nabla \cdot [P(\bm{x})\bm{v}(\bm{x})] W(\bm{x}-\bm{x}_i,h(\bm{x}))d\bm{x}\\
  &=-\int \{ \nabla\cdot [V_{\rm p}(\bm{x}) P(\bm{x})\bm{v}(\bm{x}) ] - P(\bm{x})\bm{v}(\bm{x})\cdot\nabla V_{\rm p}(\bm{x})\}  W(\bm{x}-\bm{x}_i,h(\bm{x}))d\bm{x}\\
  &=+\int  V_{\rm p}(\bm{x}) P(\bm{x})\bm{v}(\bm{x}) \cdot\nabla W(\bm{x}-\bm{x}_i,h(\bm{x})) + P(\bm{x})\bm{v}(\bm{x})\cdot\nabla V_{\rm p}(\bm{x})  W(\bm{x}-\bm{x}_i,h(\bm{x}))d\bm{x}\\
  &=+\sum_j\int P(\bm{x})\bm{v}(\bm{x})\cdot V_{\rm p}^2(\bm{x})[W(\bm{x}-\bm{x}_j,h(\bm{x}))\nabla W(\bm{x}-\bm{x}_i,h(\bm{x}))-W(\bm{x}-\bm{x}_i,h(\bm{x}))\nabla W(\bm{x}-\bm{x}_j,h(\bm{x}))] d\bm{x}\\
  &=-\sum_j\int P(\bm{x})\bm{v}(\bm{x})\cdot V_{\rm p}^2(\bm{x}) \left[ (\nabla_i-\nabla_j)-\frac{h}{2}\nabla h (\nabla_i^2 -\nabla_j^2) \right] W(\bm{x}-\bm{x}_i,h(\bm{x}))W(\bm{x}-\bm{x}_j,h(\bm{x}))d\bm{x}.
\end{align}

\subsubsection{Evaluation of convolution integrals}\label{evaluation}
Similar to the case with a fixed smoothing length, analogous integral terms appear in both equations.
The difference from the previous case lies in the derivative terms of the smoothing length and the spatial dependence of the smoothing length in the arguments of the kernel function.
Assuming that the smoothing length varies smoothly thanks to $C_{\rm smooth} > 1$ in Eqs.(\ref{eq:def_h}) and (\ref{eq:def_Vp*}), we approximate as
\begin{align}
    &\nabla h \approx 0,\\
    &(\nabla_i-\nabla_j)\int W(\bm{x}_i-\bm{x}_j,\sqrt{2}h(\bm{x})) d\bm{x} \approx \nabla_iW(\bm{x}_i-\bm{x}_j,\sqrt{2}h_i) - \nabla_jW(\bm{x}_i-\bm{x}_j,\sqrt{2}h_j),
\end{align}
where this approximation is introduced to preserve symmetry between the $i$-th and $j$-th particles.
Here, we treat the change in smoothing length from $x = x_i$ to $x = x_j$  as sufficiently small to be negligible.
From this, the integral is approximated as follows;
\begin{align}
    &\int f(\bm{x})V_{\rm p}^2(\bm{x})\left[ (\nabla_i-\nabla_j)-\frac{h}{2}\nabla h (\nabla_i^2 -\nabla_j^2) \right] W(\bm{x}-\bm{x}_i,h(\bm{x}))W(\bm{x}-\bm{x}_j,h(\bm{x}))d\bm{x}\\
    &=\int f(\bm{x})V_{\rm p}^2(\bm{x})\left[ (\nabla_i-\nabla_j)-\frac{h}{2}\nabla h (\nabla_i^2 -\nabla_j^2) \right]\left( \frac{1}{h(\bm{x})\sqrt{\pi}} \right)^{2d}\exp\left[ -\frac{ 2\left( \bm{x}-\frac{\bm{x}_i+\bm{x}_j}{2} \right)^2+\frac{(\bm{x}_i-\bm{x}_j)^2}{2} }{h^2(\bm{x})} \right]d\bm{x}\\
    &\approx f_{ij}V^2_{ij}[\nabla_iW(\bm{x}_i-\bm{x}_j,\sqrt{2}h_i) - \nabla_jW(\bm{x}_i-\bm{x}_j,\sqrt{2}h_j)],
\end{align}
where
\begin{align}
    {V}^2_{ij} &\equiv \frac{1}{2}\left( {V}^2_{ij}(h_i)+{V}^2_{ij}(h_j) \right).
\end{align}
We also use the Riemann solver $f^*_{ij}$ 
 to approximate $f_{ij}$ as before.
Therefore, SRGSPH's equations of motion and energy are yielded as follows;
\begin{align}
\label{eq:SRGSPH_EoM}
    \ev{\nu_i\dot{\bm{S}}_i} &= - \sum_j P^*_{ij}V^2_{ij,\, {\rm interp}}[\nabla_iW(\bm{x}_i-\bm{x}_j,\sqrt{2}h_i) - \nabla_jW(\bm{x}_i-\bm{x}_j,\sqrt{2}h_j)], \\
\label{eq:SRGSPH_EoE}
    \ev{\nu_i\dot{e}_i} &= - \sum_j P^*_{ij}\bm{v}^*_{ij}\cdot V^2_{ij,\, {\rm interp}}[\nabla_iW(\bm{x}_i-\bm{x}_j,\sqrt{2}h_i) - \nabla_jW(\bm{x}_i-\bm{x}_j,\sqrt{2}h_j)].
\end{align}
Eqs.(\ref{eq:SRGSPH_EoM}) and (\ref{eq:SRGSPH_EoE}) satisfy the anti-symmetry of the indices $i$ and $j$ because $P^*_{ij}=P^*_{ji}$, $\bm{v}^*_{ij}=\bm{v}^*_{ji}$, $V^2_{ij}=V^2_{ji}$, $f^*_{ij}=f^*_{ji}$ and 
The term $[ \nabla_i W(\bm{x}_i - \bm{x}_j, \sqrt{2} h_i) - \nabla_j W(\bm{x}_i - \bm{x}_j, \sqrt{2} h_j) ]$ is antisymmetric with respect to the exchange of the $i$-th and $j$-th particles
, and show that total momentum and energy are conserved.

\subsubsection{High-Accuracy Method}
To obtain $P^*_{ij}$ and $\bm{v}^*_{ij}$, we solve the Riemann problem between particle $i$ and $j$ using the primitive variables.
Minimal viscosity is introduced by solving the appropriate Riemann problem.
In this paper, we use the MUSCL method to make a spatially second-order \citep{vanLeer_1979}.
As with grid methods, these higher-order methods require monotonicity constraints to be imposed for stability.
We imposed them as follows:
\begin{align}
\label{eq:monotonicity}
    \left. \frac{\partial f}{\partial s} \right|_{\bm{x}_i}=0=\left. \frac{\partial f}{\partial s} \right|_{\bm{x}_j}\ {\rm if}\ 
    C_{\rm shock} \bm{e}_{ij}\cdot (\bm{v}_i-\bm{v}_j)>\min(c_{s,\,i},c_{s,\,j})\ \vee\ 
    \left| \log_{10}\left(\frac{P_i}{P_j}\right) \right|>C_{\rm c.d.},
\end{align}
where $f=n,\,P,\,\bm{v}$, and $\bm{e}_{ij}$ is the unit vector in the $s$-direction $\bm{e}_{ij}\equiv(\bm{x}_i-\bm{x}_j)/|\bm{x}_i-\bm{x}_j|$.
$c_s$ is the sound speed and this is derived from 
$c_s=\sqrt{(\mathit{\Gamma}-1)(H-1)/H}$.
And also $C_{\rm shock}$ and $C_{\rm c.d.}$ are numerical constants corresponding shock and contact discontinuity respectively.
 We introduce the final inequality to take care for cases with large tangential velocities. In the shock tube problem shown in the subsequent sections we find that this switch works fine.
We adopt $C_{\rm shock}=3$ and $C_{\rm c.d.}=1$ throughout in this paper.
Because the gradient terms have no impact on the conservation properties, any reasonable evaluation may be used. In this work, we use a simple formula of SPH for simplicity.
In addition to this, if the pressure or density becomes negative, we switch to the first-order method for the Riemann problem of the corresponding pair. 

\subsection{Primitive Variable Recovery}
As mentioned at the beginning of the section, we use the definition of the number density of SPH particles instead of the equation of continuity.
The time-evolving physical quantities are $\ev{\bm{S}_i}$ and $\ev{e_i}$.
From these two physical quantities, it is necessary to recover primitive physical quantities such as velocity, position or thermodynamic quantities, but the relation is complicated and can not be represented algebraically.

From Eqs.(\ref{eq:S}) to (\ref{eq:EoS}), we obtain a quartic equation of $\ev{\gamma_i}$ as follows;
\begin{align}
    (\ev{\gamma_i}^2-1)(X\ev{e_i}\ev{\gamma_i}-1)^2-\ev{\bm{S}_i}^2(X\ev{\gamma_i}^2-1)^2=0,
\end{align}
where
\begin{align}
    X\equiv \frac{\gamma_c}{\gamma_c-1}.
\end{align}
Once $\ev{\gamma_i}$ is obtained, $\ev{\bm{v}_i}$ is obtained as
\begin{align}
    \ev{\bm{v}_i}=\frac{X\ev{\gamma_i}^2-1}{\ev{\gamma_i}(X\ev{e_i}\ev{\gamma_i}-1)}\ev{\bm{S}_i}.
\end{align}
The position of the SPH particle at the next time step is derived by the time integration and the other primitive physical quantities are obtained from this velocity equation.

\section{Numerical Tests}
From this section, we will choose the coordinate system $c=1$.
We test our method with variable smoothing length on some problems.
We use Exact Riemann solver for $f^*_{ij}$ \citep{Pons_etal_2000}.
The Euler method is used as a time integration method;
\begin{align}
  \ev{\nu_i\bm{S}_i}^{n+1} &= \ev{\nu_i\bm{S}_i}^n + \ev{\nu_i\dot{\bm{S}}_i}\Delta t,\\
  \ev{\nu_ie_i}^{n+1} &= \ev{\nu_ie_i}^n + \ev{\nu_i\dot{e}_i}\Delta t,\\
  \bm{x}_i^{n+1} &= \bm{x}_i^{n} + \ev{\bm{v}_i}\Delta t.
\end{align}
In the test calculations presented in this paper, we find that a simpler time step criterion, similar to that used in non-relativistic cases, works well without than the more elaborate methods employed in \citep{Chow_Monaghan_1997} and \citep{Rosswog_2010}.
Specifically, we adopt the following Courant condition;
\begin{align}
  \Delta t = C_{\rm CFL}\min_i \left[\frac{h_i}{c_{s,i}}\right],
\end{align}
where $C_{\rm CFL}$ is a Courant constant.
We use $C_{\rm CFL} = 0.3$.

For parallelization, the Framework for Developing Particle Simulators \citep{Iwasawa+_2016} is utilized.

\subsection{Riemann Problems}
In this section, we show some results of Riemann problems.
The first two test problems correspond to one-dimensional problems without tangential velocity.
A Riemann problem with tangential velocity is considered in Sec. \ref{sec:result-withVt}. 

\subsubsection{Riemann Problem 1: Sod Problem}\label{1Dsod}
\begin{figure}[htbp]
  \begin{center}
   \includegraphics[width=100mm]{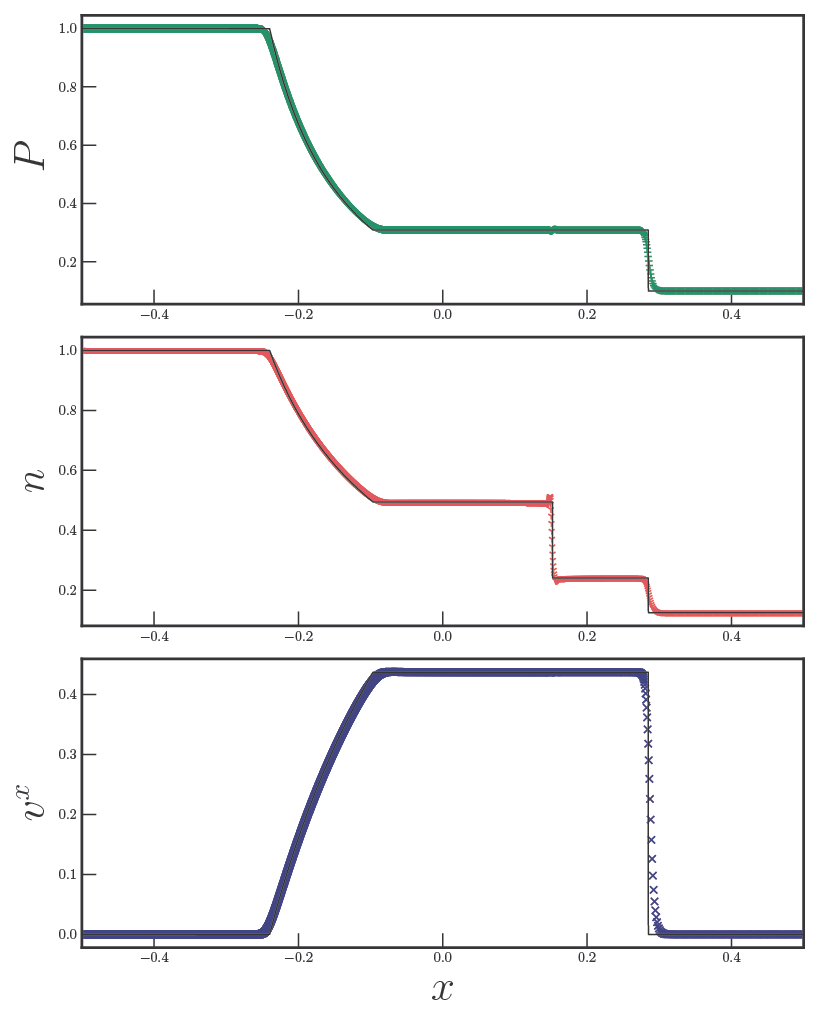}
  \end{center}
   \caption{
   The result of the Sod problem at $t = 0.35$. The symbols represent the numerical solutions and the solid lines show the analytical solutions. For this problem, each SPH particle has the same baryon number. The left-hand side contains 3200 SPH particles and the right-hand side contains 400. Calculated with $\gamma_c = 5/3$.
   \label{fig:sod}
   }
\end{figure}

To evaluate our method, we test the standard problem known as Sod problem.
In this test, the left- and right-hand physical quantities were set as follows:
\begin{align}
  (P_L, n_L, v^x_L, v^t_L) &= (1.0,\ 1.0,\ 0,\ 0),\\
  (P_R, n_R, v^x_R, v^t_R) &= (0.1,\ 0.125,\ 0,\ 0),
\end{align}
where $v^t_L$ and $v^t_R$ are the velocity components along the partition in the left and right regions, respectively.
The gas is considered as an ideal gas with the heat capacity ratio $\gamma_c = 5/3$.
The initial discontinuity is set at $x=0$.
Using SPH particles that have equal baryon numbers,
3200 particles are set on the left-hand side of the initial discontinuity and 400 particles on the right-hand side.
The result at $t=0.35$ are shown in Fig.\ref{fig:sod}.
In most regions of the result, they are consistent with the analytical solution.
This slight pressure fluctuations at the contact discontinuity and this is a common issue in many SPH methods, regardless of whether it is relativistic or non-relativistic, but the amplitude of this pressure wiggle in Godunov SPH is much smaller than that of standard SPH.
The pressure error at the contact discontinuity in the subsequent test calculations is similar.

\begin{figure}[htbp]
  \begin{center}
    \includegraphics[width=75mm]{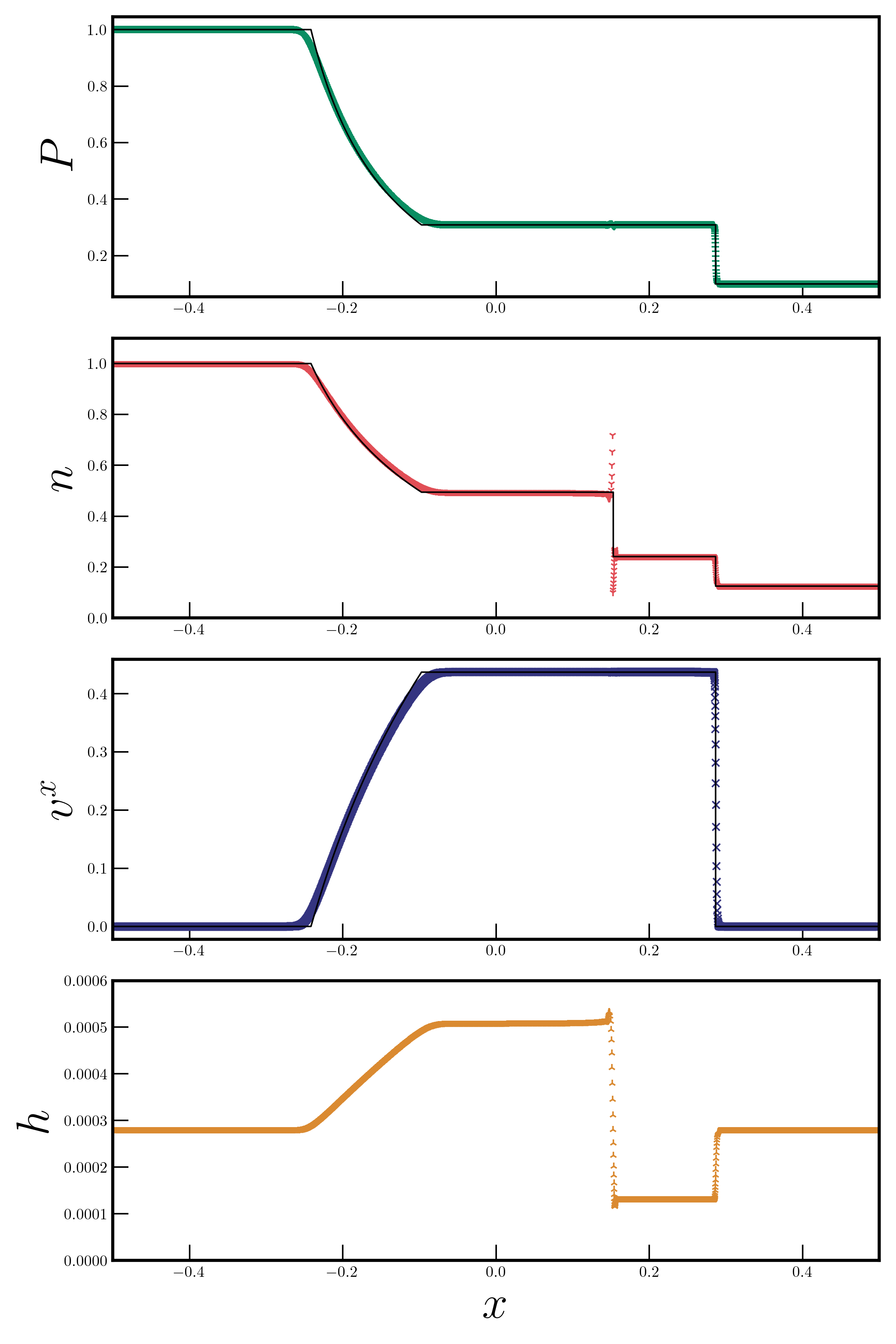}
    \includegraphics[width=75mm]{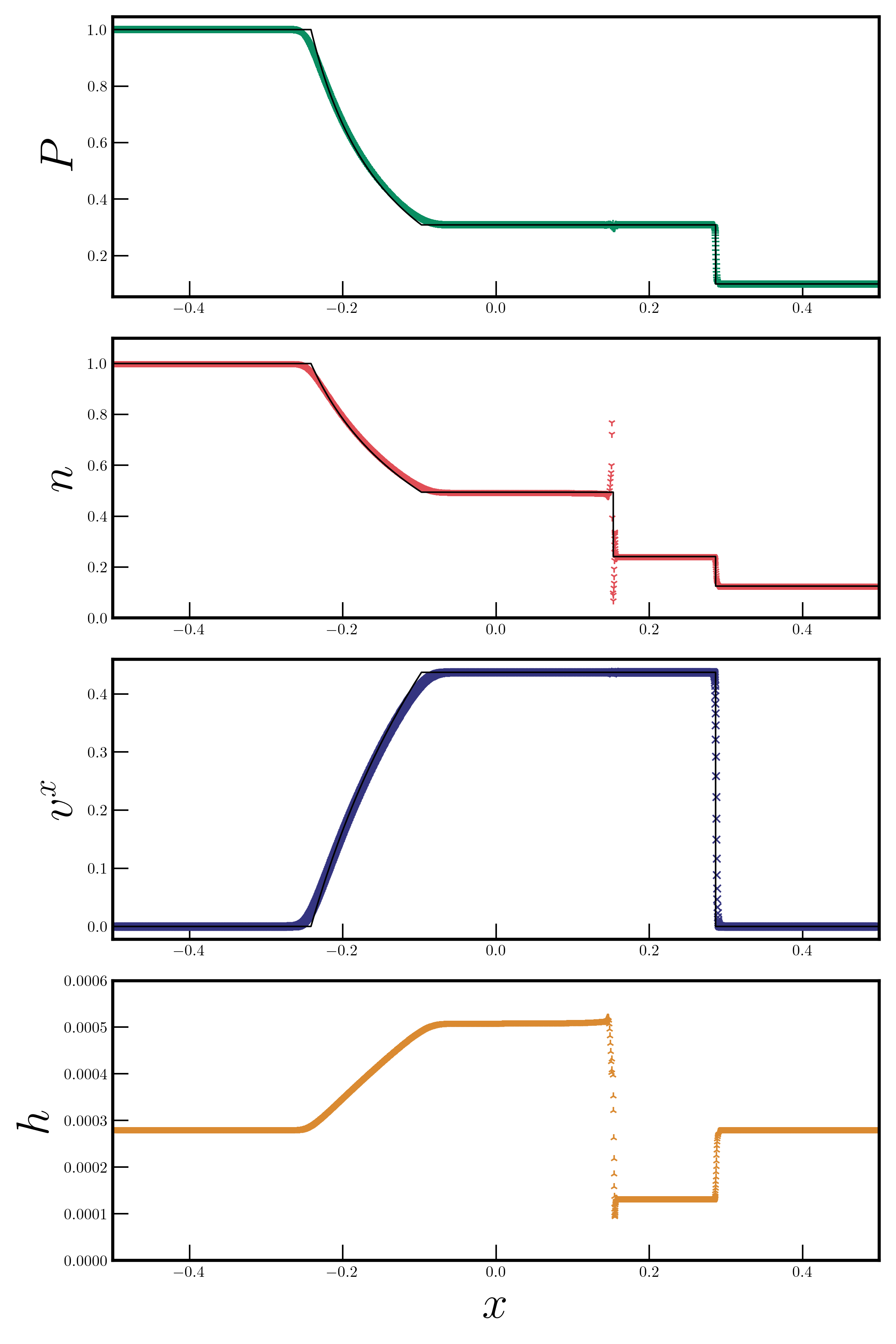}
  \end{center}
  \caption{
    Results of the Sod problem at $t = 0.35$ with different baryon numbers across the initial discontinuity.
    Both panels show numerical solutions (symbols) compared with the analytical solutions (solid lines).
    The left panel corresponds to the volume-based approach, and the right panel to the standard approach.
    The number of SPH particles on each side of the discontinuity is 1800.
    We set $\gamma_c = 5/3$.
  }
  \label{fig:sod2}
\end{figure}

Next, to evaluate the volume-based approach, we perform a test in which the baryon numbers differ between particles.
Fig. \ref{fig:sod2} presents the results obtained with 1800 particles placed on both sides of the initial discontinuity.
Although a density jump appears at the contact discontinuity, the smoothing length profile connects smoothly across the discontinuity.
This behavior arises for the reason illustrated in Fig.\ref{fig:Vb_vs_Stn_2}.
Comparing the volume-based approach (left panel in Fig. \ref{fig:sod2}) with the standard approach (right panel in Fig. \ref{fig:sod2}), the volume-based method exhibits smaller overshoot and undershoot than the standard approach.
This behavior not only improves the physical consistency of the method but is also advantageous from a computational perspective, since milder deviations in the smoothing length contribute to more stable and efficient time integration.

\subsubsection{Riemann Problem 2: Standard Relativistic Blast Wave Problem}
\begin{figure}[htbp]
  \begin{center}
   \includegraphics[width=100mm]{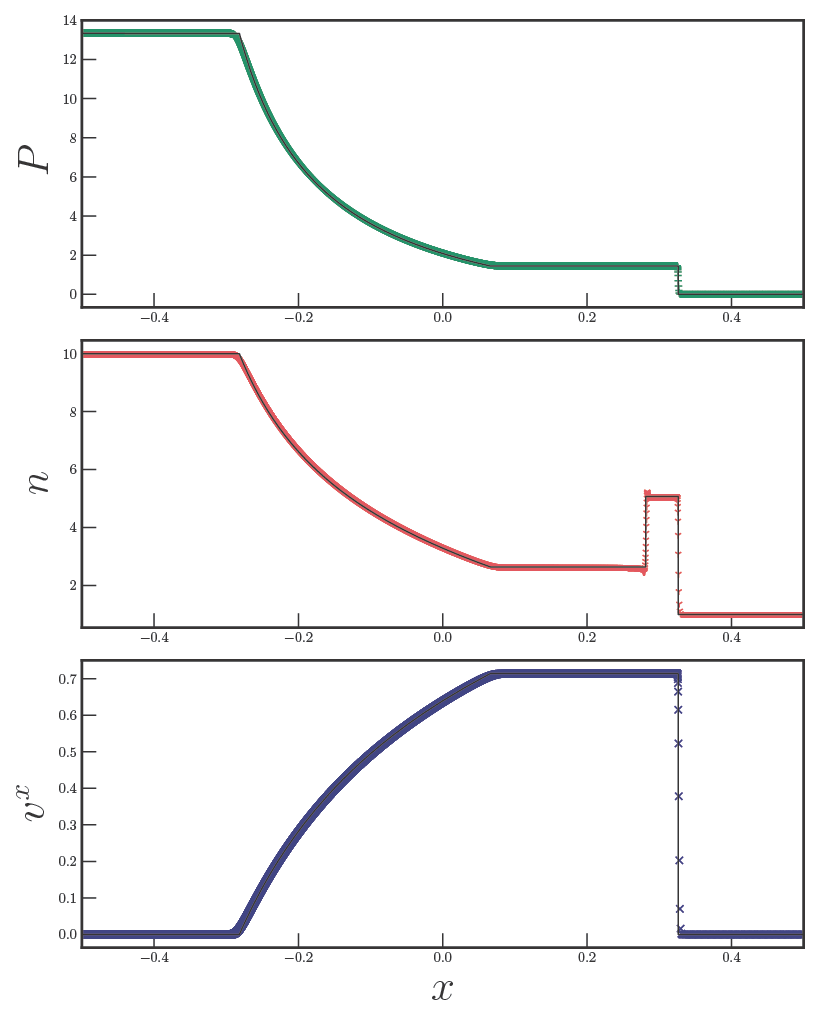}
  \end{center}
   \caption{
   The result of the standard relativistic blast wave problem at $t = 0.4$. The symbols represent the numerical solutions and the solid lines show the analytical solutions. For this problem, SPH particles have equal baryon numbers. The left-hand side contains 5000 SPH particles and the right-hand side contains 500. Calculated with $\gamma_c = 5/3$.
   \label{fig:stn}
   }
\end{figure}

In the Riemann problem with a large initial pressure jump, a blast wave occurs with a high-density shell propagating at relativistic speeds; see \citep{Marti&Muller_2003} for details.
In this section, we calculate a relatively simpler test problem compared to the next section.
In this test, the left- and right-hand physical quantities were set as follows:
\begin{align}
  (P_L, n_L, v^x_L, v^t_L) &= (40.0/3.0,\ 10.0,\ 0,\ 0),\\
  (P_R, n_R, v^x_R, v^t_R) &= (10^{-6},\ 1.0,\ 0,\ 0).
\end{align}
We use the heat capacity ratio $\gamma_c = 5/3$.
Using SPH particles with equal baryon numbers,
5000 particles are set on the left-hand side of the initial discontinuity and 500 particles on the right-hand side.
The results at $t=0.4$ are shown in Fig.\ref{fig:stn}.

\begin{figure}[htbp]
  \begin{center}
    \includegraphics[width=75mm]{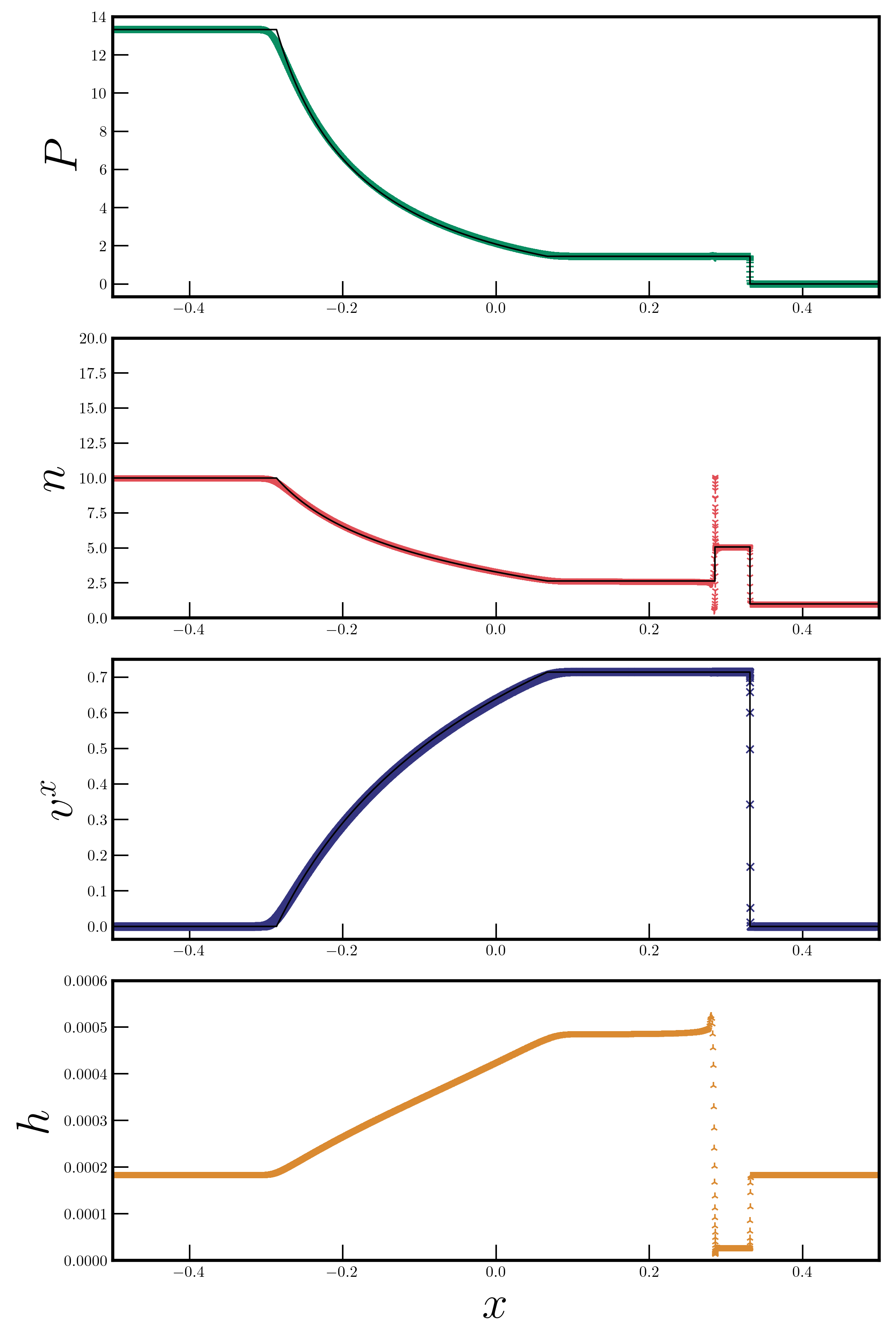}
    \includegraphics[width=75mm]{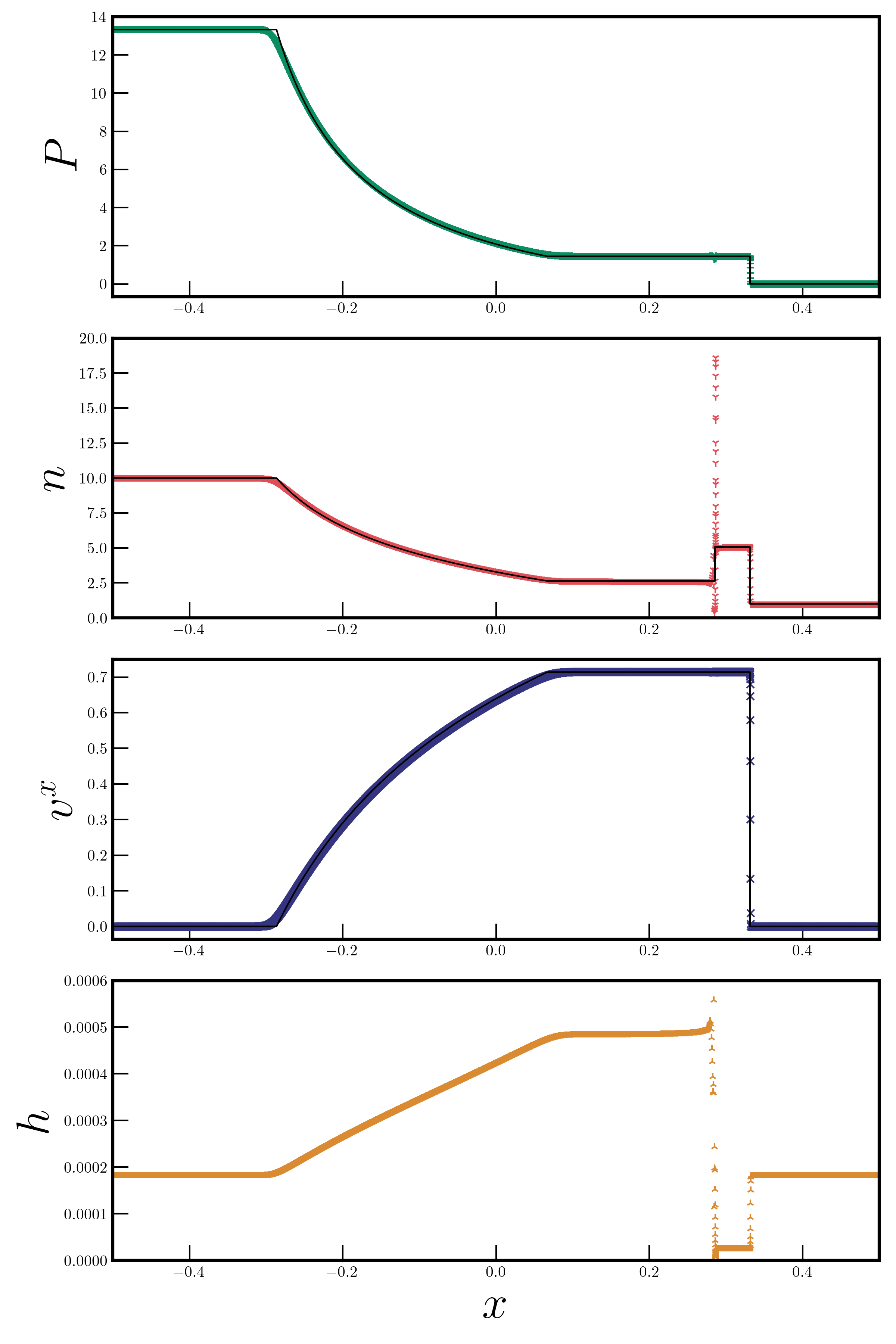}
  \end{center}
  \caption{
  Results of the standard relativistic blast wave problem at $t = 0.4$, where the baryon number in each particle on the left hand side is uniform but its value is different from that on the right hand side.  
  The number of SPH particles on both sides of the initial discontinuity is 2750.
    The left panel corresponds to the volume-based approach, and the right panel to the standard approach.
  The calculation is performed with $\gamma_c = 5/3$.
  \label{fig:stn2}
  }
\end{figure}

As in the previous test, we next examine the case where the baryon numbers differ between particles.
Fig. \ref{fig:stn2} shows the result when 2750 particles are placed on both the left and right sides of the initial discontinuity.
The density jumps appear for the same reason as Fig. \ref{fig:sod2}.

\subsubsection{Riemann Problem 3: Strong Relativistic Blast Wave Problem}
\begin{figure}[htbp]
  \begin{center}
   \includegraphics[width=120mm]{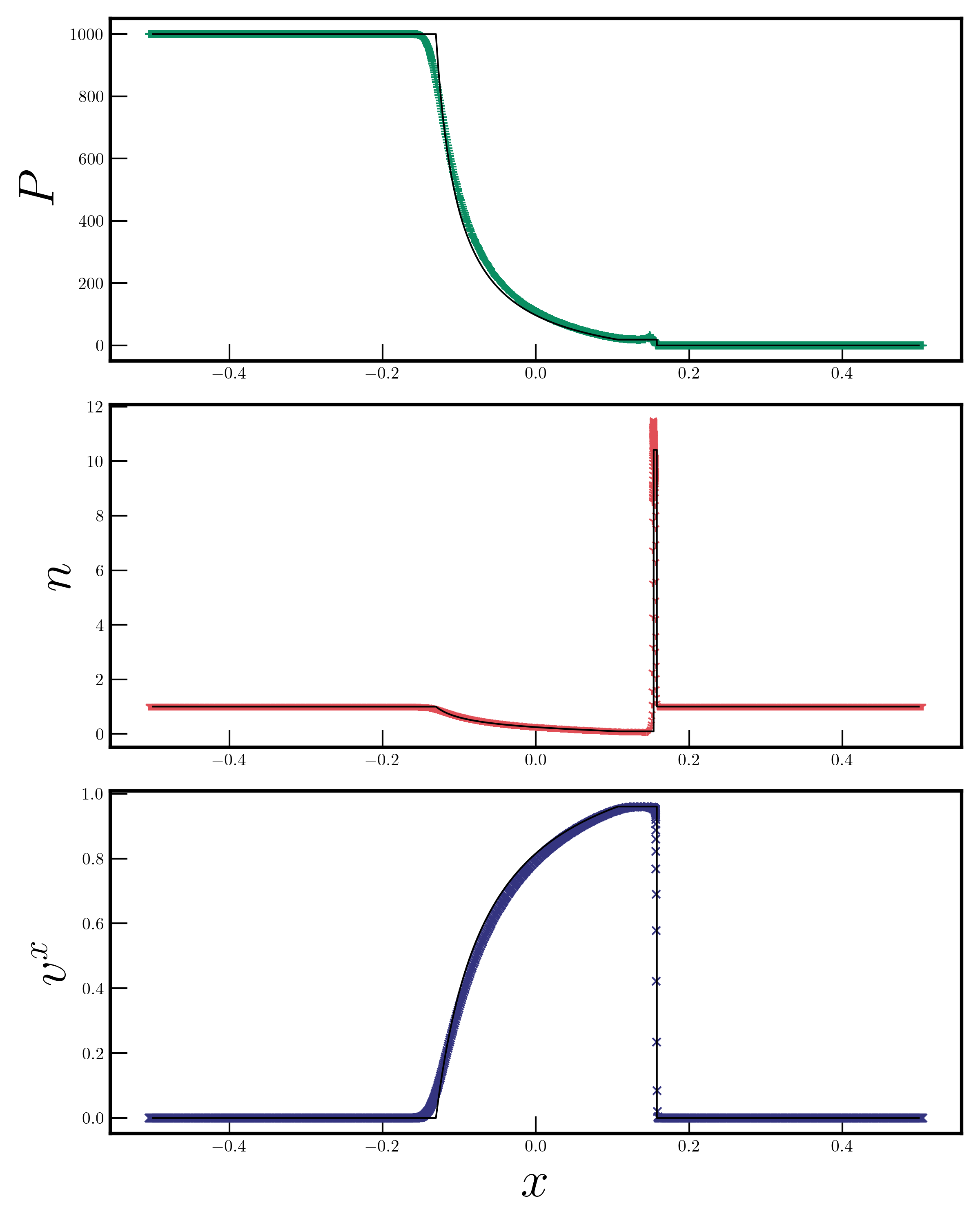}
  \end{center}
   \caption{
   The result of the strong relativistic blast wave problem at $t = 0.16$. The symbols represent the numerical solutions and the solid lines show the analytical solutions. For this problem, SPH particles are equal baryon numbers. Each side contains 900 SPH particles. Calculated with $\gamma_c = 5/3$.
   \label{fig:strong}
   }
\end{figure}

A stronger relativistic blast wave problem is considered in this section, extending the test performed in the previous one.
This test has been studied in previous works using both grid-based schemes \citep[e.g.,][]{Marti&Muller_2003} and SPH methods \citep[e.g.,][]{Rosswog_2010,Liptai&Price_2019}, providing a useful benchmark for comparison.
In this test, the left- and right-hand physical quantities were set as follows:
\begin{align}
  (P_L, n_L, v^x_L, v^t_L) &= (1000.0,\ 1.0,\ 0,\ 0),\\
  (P_R, n_R, v^x_R, v^t_R) &= (0.01,\ 1.0,\ 0,\ 0).
\end{align}
We use the heat capacity ratio $\gamma_c = 5/3$.
Using SPH particles with equal baryon numbers,
900 particles are set on each side of the initial discontinuity.
The results at $t=0.16$ are shown in Fig.\ref{fig:strong}.

\begin{figure}[htbp]
  \begin{center}
   \includegraphics[width=120mm]{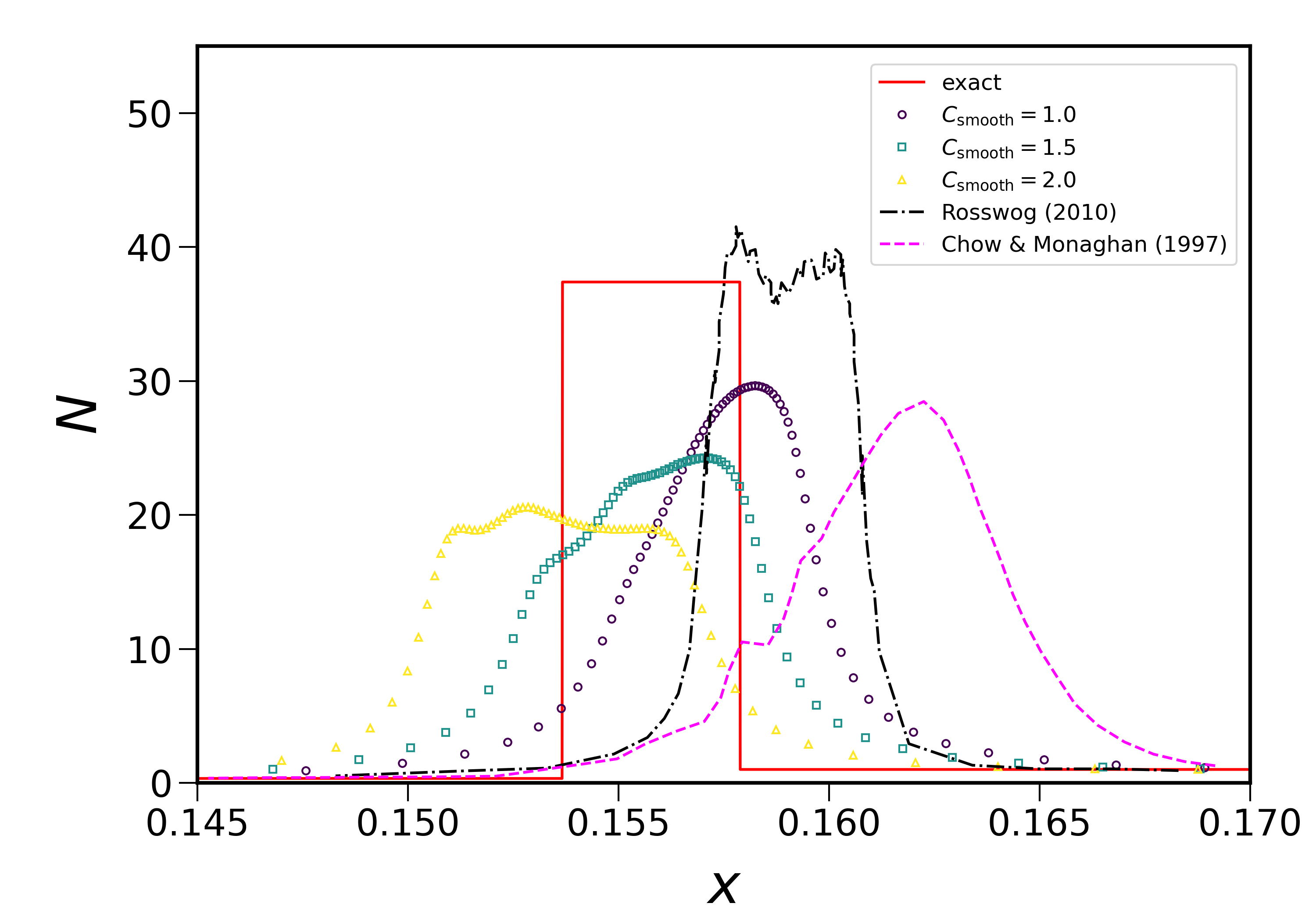}
  \end{center}
   \caption{
   Number density profiles in the laboratory frame, focusing on the thin shell structure formed in the strong relativistic blast wave problem.
The results are shown for various values of the smoothing coefficient $C_{\text{smooth}}$ and are compared with those from previous studies \citep{Chow_Monaghan_1997, Rosswog_2010}.
   \label{fig:strong_zoom}
   }
\end{figure}

As shown in the number density profile, a very thin layer is formed in this problem.
We therefore focus on this layer and investigate the effect of varying $C_{\text{smooth}}$.
To directly compare with previous studies \citep{Chow_Monaghan_1997, Rosswog_2010}, we also performed simulations at a reduced resolution, placing 200 particles on each side of the initial discontinuity.
In these simulations, we systematically varied $C_{\text{smooth}}$ to evaluate its impact on the solution.
The results are presented in Fig.~\ref{fig:strong_zoom}. As expected, increasing $C_{\text{smooth}}$ leads to a smoother number density profile.
We also observe that the shock front is slightly less sharp than in the reference solution of \citet{Rosswog_2010}.
This difference is attributed to the strict monotonicity condition imposed by our limiter, which suppresses steep gradients, including those at shock fronts, resulting in a smoother profile.
The relationship between $C_{\text{smooth}}$, the positions of the contact discontinuity, and the shock front is subtle.
It depends not only on the value of $C_{\text{smooth}}$ itself but also on the monotonicity condition Eq.\ref{eq:monotonicity} and the specific problem setup.
While a larger $C_{\text{smooth}}$ enhances smoothness, it can also slightly shift the positions of discontinuities.
In this calculation, we find that $C_{\text{smooth}} \approx 1.5$\text{--}$2.0$ strikes a good balance between smoothness and accuracy.

\subsection{ Riemann Problem 4: Ultra-Relativistic Shock}
\begin{figure}[htbp]
  \begin{center}
   \includegraphics[width=120mm]{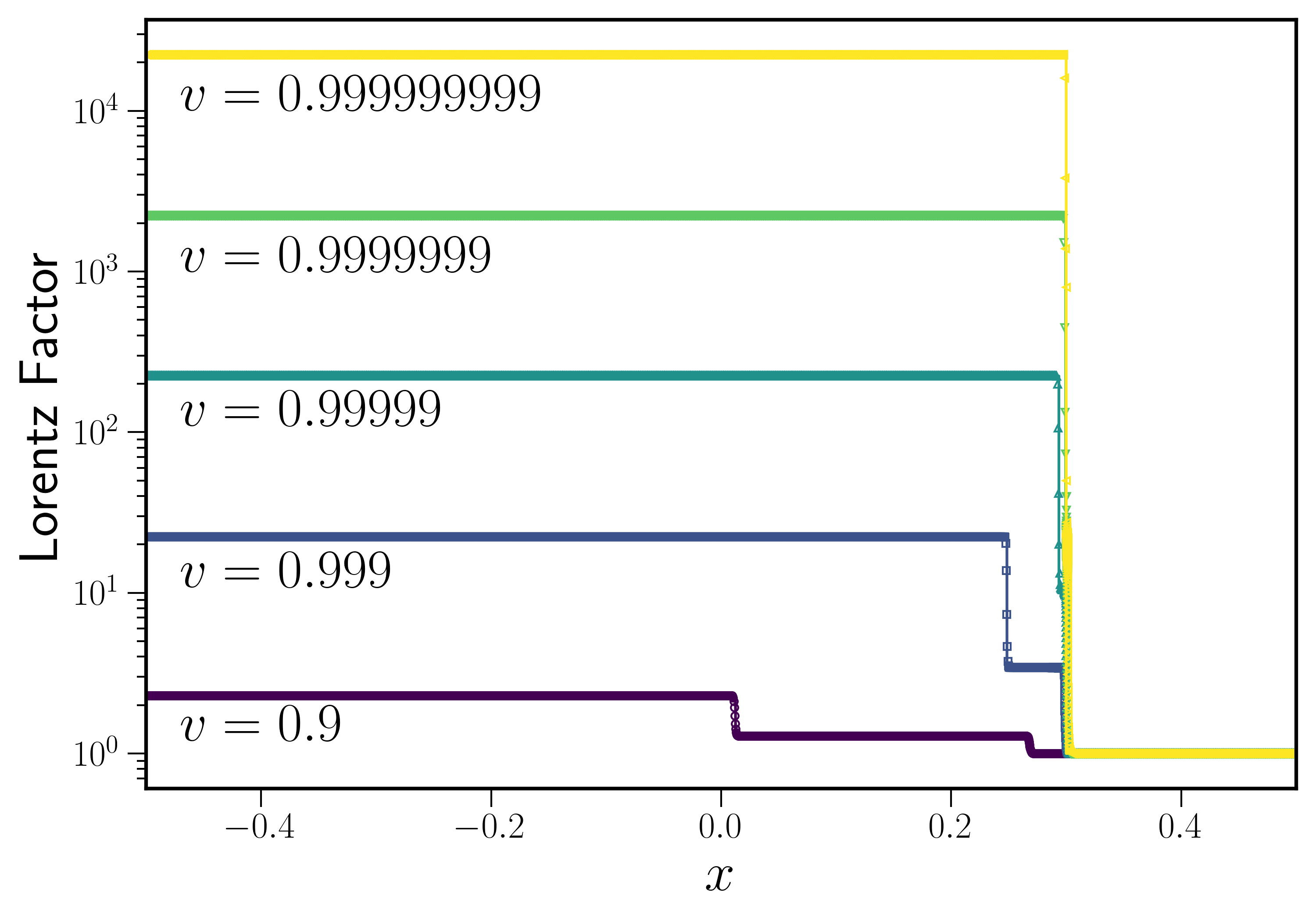}
  \end{center}
   \caption{
    Simulation results at t = 0.3 for a relativistic flow with various initial velocities. The profiles demonstrate the formation of extremely strong shock waves, confirming the robustness of the present method.
    For this test problem, both the left- and right-hand sides contain 1000 particles and $\mathit{\gamma_c}=5/3$.
   \label{fig:ult}
   }
\end{figure}

In the following test, we consider a scenario in which a flow with a high Lorentz factor impinges on a stationary fluid.
This setup allows us to assess whether the simulation code can reliably handle extremely strong shocks that emerge from such collisions.
The initial conditions are set as follows:
\begin{align}
  (P_L, n_L, v^x_L, v^t_L) &= (1.0,\ 1.0,\ \{0.9\mathchar`-0.999999999\},\ 0),\\
  (P_R, n_R, v^x_R, v^t_R) &= (1.0,\ 1.0,\ 0,\ 0).
\end{align}
Specifically, simulations are performed for five discrete values of $v^x_L$ within this range.
Fig. \ref{fig:ult} presents the simulation results at $t = 0.3$.

\subsubsection{Riemann Problem 5: One-Dimensional Shock Tube Problem With Tangential Velocity}
\label{sec:result-withVt}
\begin{figure}[htbp]
  \begin{center}
   \includegraphics[width=150mm]{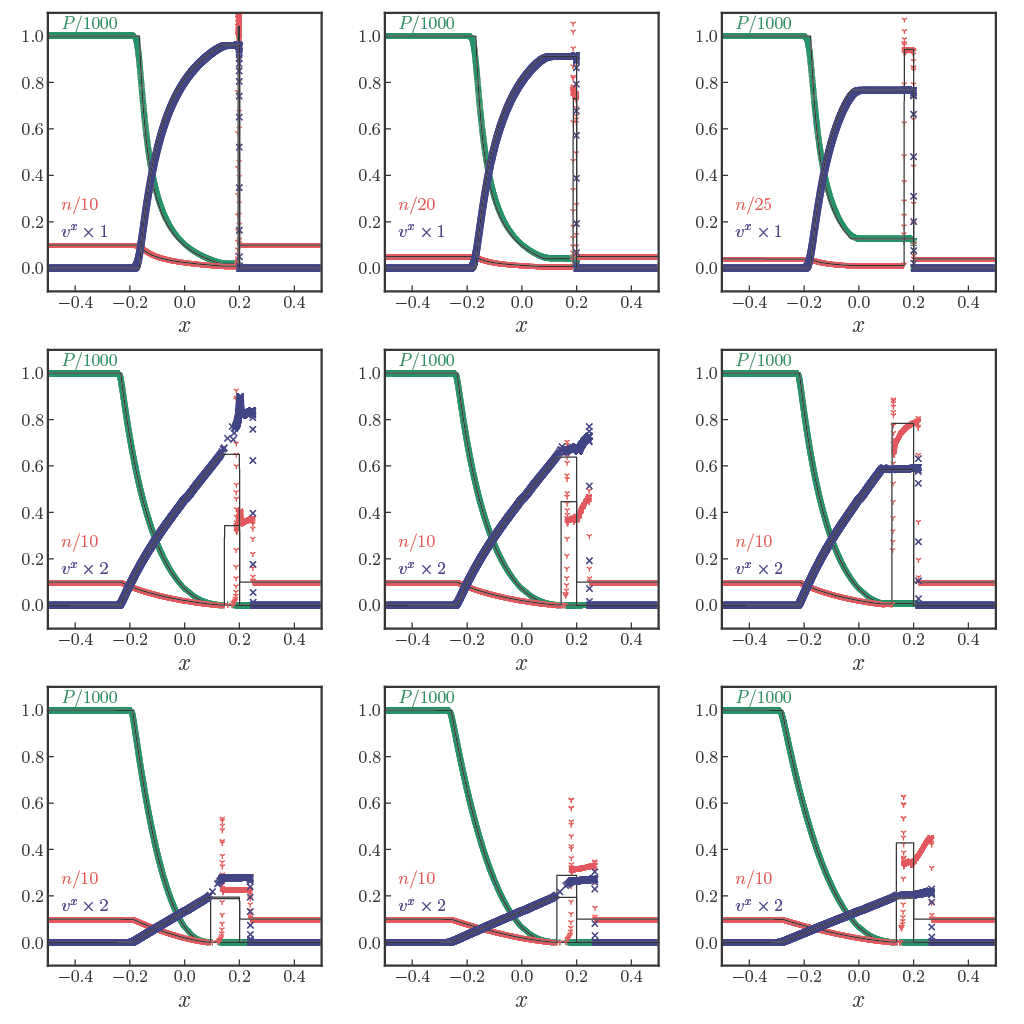}
  \end{center}
   \caption{
   The calculation result of the multidimensional special relativistic shock tube problem at the shock front reaches $x=0.2$; from left to right, $v^t_R=0,\ 0.9,\ 0.99$, and from top to bottom, $v^t_L=0,\ 0.9,\ 0.99$.
For this test problem, both the left- and right-hand sides contain 1600 particles and $\mathit{\gamma_c}=5/3$.
   \label{fig:9}
   }
\end{figure}

\begin{figure}[htbp]
  \begin{center}
   \includegraphics[width=150mm]{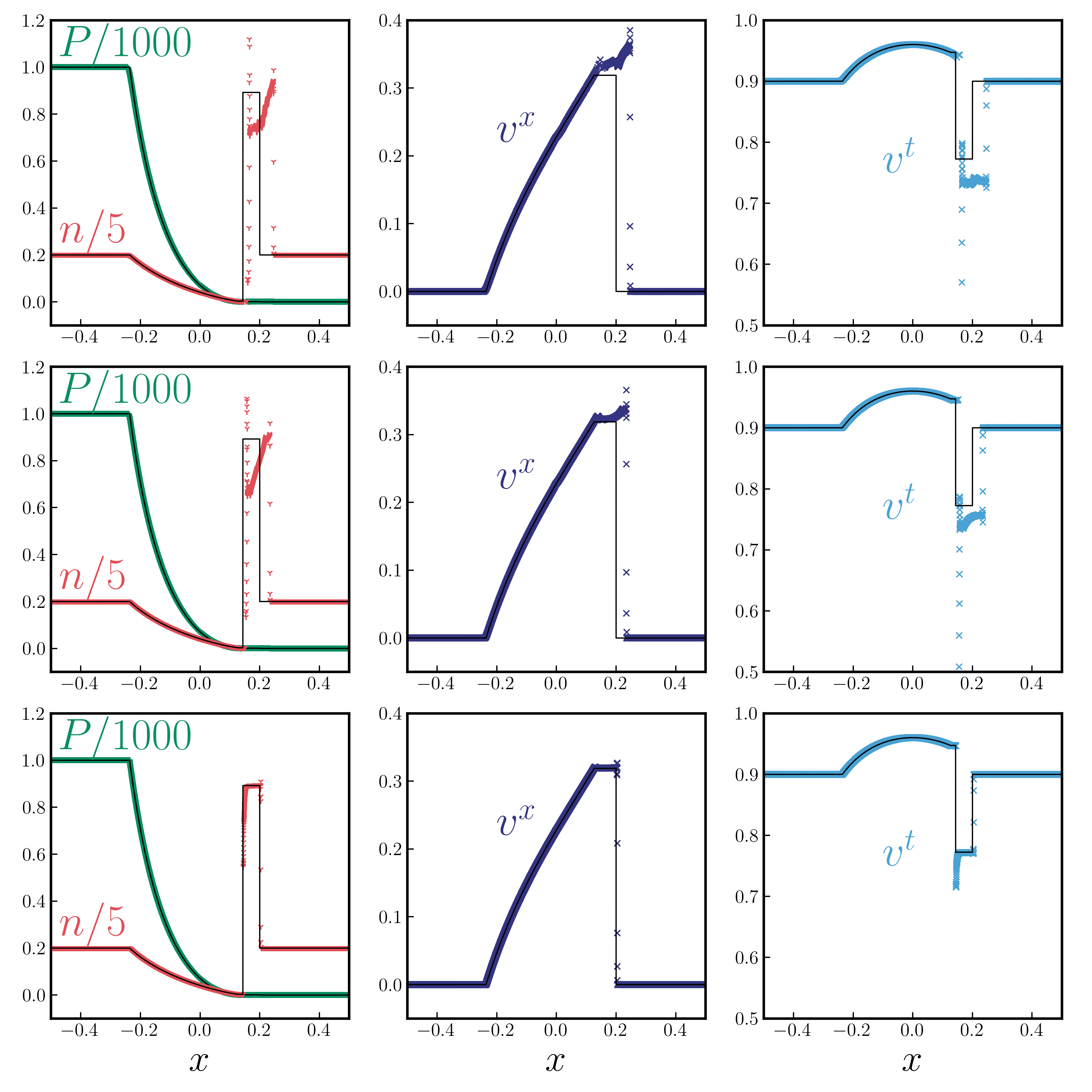}
  \end{center}
   \caption{
    The calculation result of the multidimensional special relativistic shock tube problem with $v^x_L=0.9$ and $v^t_R=0.9$.
    The number of particles on the right-hand side was fixed at 1,600, while the number of particles on the left-hand side varied from 1,600, 3,200, 51,200.
   \label{fig:3}
   }
\end{figure}

In the relativistic case, the velocity component along the partition $v^t$ affects the solution, which is in contrast to non-relativistic cases \citep[e.g.,][]{Pons_etal_2000,Rezzolla_Zanotti_2002}.
When the initial tangential velocity is large, it is known that errors in numerical solution become large and an extremely high resolution is required to obtain an accurate result \citep[e.g.,][]{Mignone_etal_2005,Zhang_MacFadyen_2006}.
Following \citet{Pons_etal_2000}, the initial conditions are set as follows:
\begin{align}
  (P_L, n_L, v^x_L, v^t_L) &= (1000.0,\ 1.0,\ 0,\ \{0,0.9,0.99\}),\\
  (P_R, n_R, v^x_R, v^t_R) &= (0.01,\ 1.0,\ 0,\ \{0,0.9,0.99\}).
\end{align}
Here, $v^t_L$ and $v^t_R$ are assigned to be either 0, 0,9 or 0.99, resulting in a total of 9 test calculations.
For this test problem, both the left- and right-hand sides contain 1600 particles.
The resulting snapshots  of the calculations when the shock front reaches $x = 0.2$ are shown in Fig. \ref{fig:9}.
As can be seen from the results, the deviation from the analytical solution increases when the initial tangential velocity is large on the rarefaction wave side.
So we focus on the resolution dependence on the rarefaction wave.
We set $v^x_L=0.9$ and $v^t_R=0.9$, and particles on the right-hand side is fixed at 1,600, while the number of particles on the left-hand side varied from 1,600, 3,200, to 51,200.
The snapshots of the calculations when the shock front reaches $x = 0.2$ are shown in Fig. \ref{fig:3}.
As can be seen in Fig. \ref{fig:3}, the increased resolution on the expansion wave side effectively resolves this issue.
However, as noted in \citep{Liptai&Price_2019}, the behavior of the tangential velocity $v^t$ near the contact discontinuity at low resolution is worse in SPH-based methods compared to grid-based methods, such as those presented in \citep{Zhang_MacFadyen_2006}.
This is likely due to the nature of SPH, in which low-density regions are inherently represented with lower resolution.

\subsubsection{Riemann Problem 6: Two-Dimensional Sod Problem}
\begin{figure}[htbp]
  \begin{center}
   \includegraphics[width=100mm]{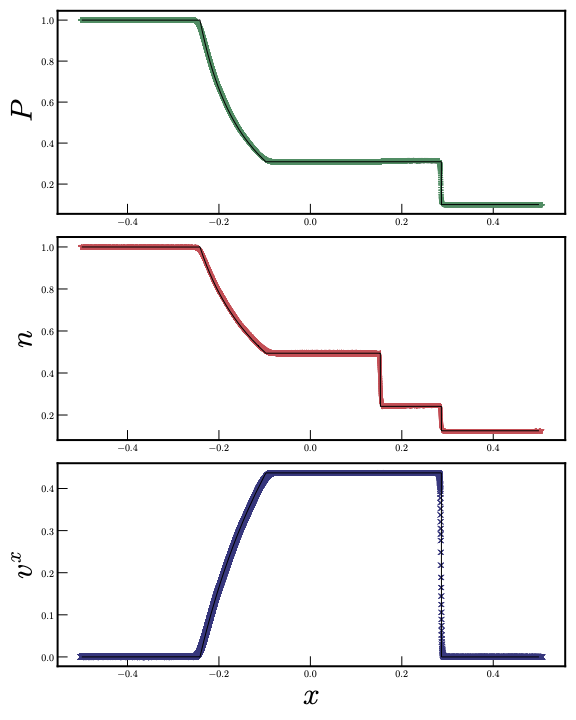}
  \end{center}
   \caption{
     The two dimensional calculation result of the Sod shock-tube problem at $t = 0.35$. For this problem, SPH particles have equal baryon numbers. The simulation is performed with 61,554 particles, with $\gamma_c = 5/3$.
   \label{fig:2Dsod}
   }
\end{figure}

Next, to evaluate the multidimensional computation, we solve the Riemann problem in two dimensions.
The initial condition is set in the same manner as described in Sec. \ref{1Dsod}, while performed with approximately 60,000, particles and periodic boundary conditions are applied in the $y$-direction. 
The particles were arranged in an equilateral triangular lattice. 
The results at $t=0.35$ are shown in Fig \ref{fig:2Dsod}, which demonstrate that the multidimensional computation can be performed with similar accuracy as in the one-dimensional case.

\subsection{Kelvin-Helmholtz Instability}
\begin{figure}[htbp]
    \begin{tabular}{cc}
      \begin{minipage}[t]{0.45\hsize}
        \centering
        \includegraphics[width=70mm]{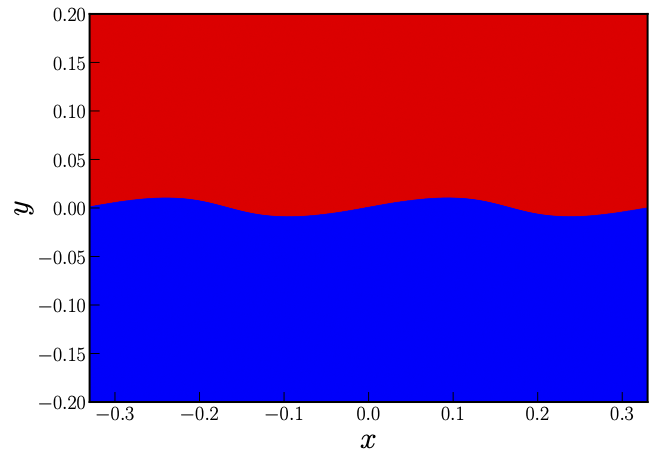}
      \end{minipage} &
      \begin{minipage}[t]{0.45\hsize}
        \centering
        \includegraphics[width=70mm]{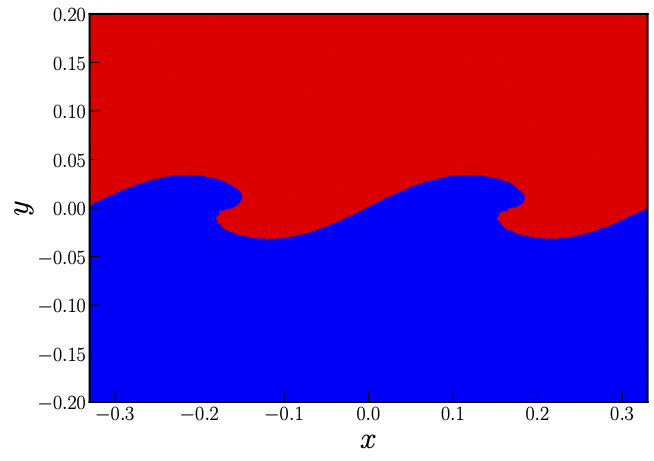}
      \end{minipage} \\
   
      \begin{minipage}[t]{0.45\hsize}
        \centering
        \includegraphics[width=70mm]{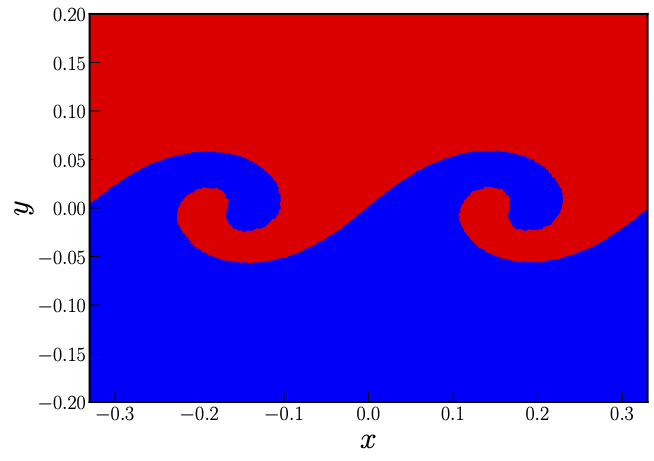}
      \end{minipage} &
      \begin{minipage}[t]{0.45\hsize}
        \centering
        \includegraphics[width=70mm]{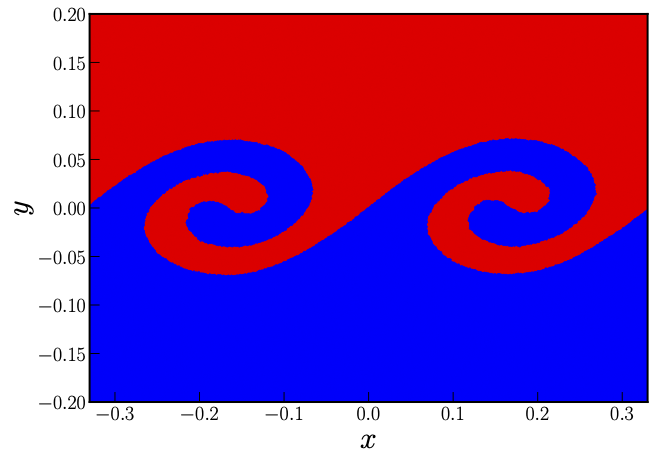}
      \end{minipage} 
    \end{tabular}
     \caption{The results of two-dimensional Kelvin-Helmholtz instability simulation at $t=0.5,\ 1.0,\ 1.5,\ 2.0$ are shown. The colors represent the initial positions: red indicates $y>0$, while blue indicates $y<0$. The simulation is performed with approximately 250,000 particles. Calculated with $\gamma_c = 5/3$.
   \label{fig:KH}}
  \end{figure}

\begin{figure}[htbp]
  \begin{center}
   \includegraphics[width=100mm]{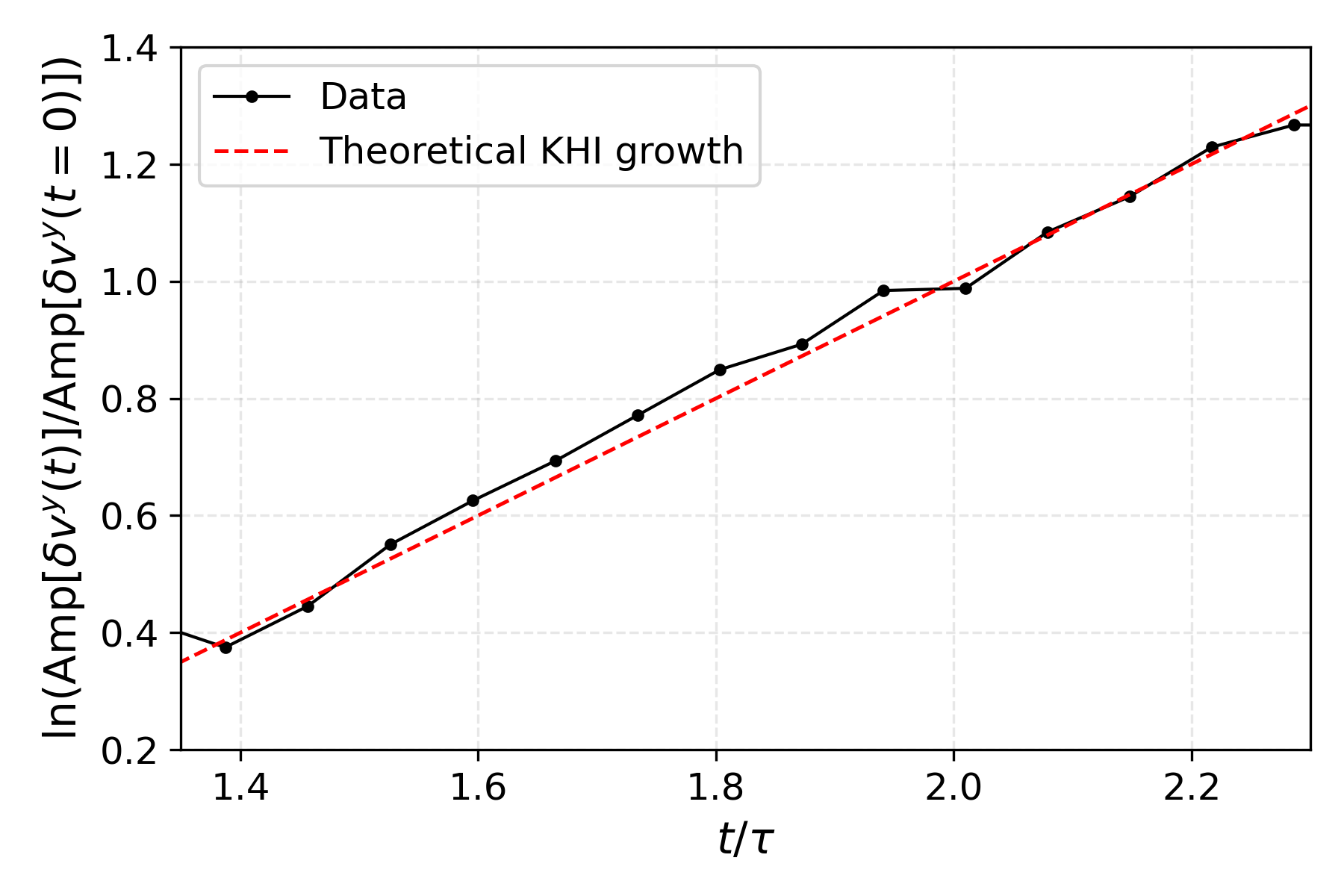}
  \end{center}
   \caption{
    Time evolution of the amplitude of the $v^y$. 
The growth rate is measured during the linear phase of the instability and compared with the theoretical prediction by \citet{Bodo&Rosner_2004}.
Here, $\tau$ denotes the characteristic growth time, defined as the inverse of the growth rate obtained by the linear analysis, and is approximately $0.354$ for the present setup. $\mathrm{Amp}[\delta v^y(t=0)]$ represents the initial perturbation amplitude, i.e., $A_0$.
   \label{fig:KHgrow}
   }
\end{figure}

We also test our method for Kelvin-Helmholtz Instability in the framework of special relativity. 
The initial condition is a steady state where two layers are moving at different velocities.
Pressure is set to 1, and densities of both the upper and lower layers are set to 0.5. 
The two layers move in opposite directions with a velocity of 0.3.
An initial perturbation is introduced in the $y$-direction as follows: 
\begin{align}
    v^y_{1}(x,y) = A_0 \sin \left( 2\pi\frac{x}{\lambda} \right),\ {\rm if}\ |y|<0.05
\end{align}
where 
$A_0$ represents the amplitude of the perturbation and is set to 1/40, while 
$\lambda$ represents the wavelength of the initial perturbation and is fixed at 1/3.
This perturbation is applied only to a thin layer surrounding the contact discontinuity (where $|y|<0.05$).
Periodic boundary conditions are applied in the $x$-direction, and reflective boundary conditions are used in the $y$-direction.
The total number of particles in the computational domain is approximately 250,000.
The results are shown in Fig. \ref{fig:KH}, which demonstrate the capability of the present method.
As Fig. \ref{fig:KHgrow}, this result is consistent with the growth rate predicted in \citep{Bodo&Rosner_2004}, indicating that the present method accurately captures, at least, the linear phase of the Kelvin-Helmholtz instability in the relativistic regime.
The ability of the instability to grow without thermal conduction can be understood in light of findings in the non-relativistic regime, where convolution-based formulations have been shown to reduce numerical dissipation across contact discontinuities more effectively than conventional methods \citep{Cha+2010}.
The current simulation may thus be regarded as a relativistic extension of such approaches.
Furthermore, the volume-based formulation employed in this work enables precise control over the smoothing length by adjusting the baryon number, which ensures that the smoothing lengths on both sides of the interface remain consistent.
This feature enhances the susceptibility of the system to the development of the instability.
The morphology and growth pattern of the instability observed in our simulation are consistent with other simulations of Kelvin-Helmholtz Instability without physical viscosity \citep[e.g.,][]{Duffell+2011}.
The effect of the physical and numerical viscosity is discussed in \citet{Townsend+2022}.

\section{DISCUSSION \& SUMMARY}
We have developed an extension of Godunov SPH to perform numerical simulations of relativistic hydrodynamics.
By using convolution integrals to define the physical quantities of SPH particles, the formulation achieves higher accuracy compared to conventional methods.
However, as described in Sec. \ref{evaluation_const}, the computational cost associated with numerical integration is avoided by analytically interpolating the number density, allowing the convolution integrals to be evaluated in closed form.
While this approximation introduces some pressure errors at contact discontinuities, these errors are significantly smaller than those observed in standard SPH formulations.
Additionally, by defining particle volume as the field quantity, the spatial change of the smoothing length remains monotonic around a contact discontinuity even for SPH particles with different baryon numbers. Under this framework, the physical viscosity to describe shock waves is introduced implicitly through the use of a Riemann solver.
By reformulating the equations to preserve index symmetry—ensuring that the action-reaction principle holds—the scheme conserves total momentum and energy.
This conservation property is maintained even when the smoothing length is allowed to vary.
The performance of SRGSPH is validated through a series of numerical tests, including one- and two-dimensional shock tube problems and simulations of the Kelvin-Helmholtz instability. These tests demonstrate the method’s robustness, accuracy, and applicability to high-resolution simulations. By enabling efficient handling of complex relativistic flows and providing a more accurate treatment of density variations, SRGSPH offers a significant advancement in SPH methods, with promising implications for further applications in high-energy astrophysical simulations.
An extension of the present method to general relativistic hydrodynamics, along the lines of \citep[e.g.,][]{Rosswog_2010GR,Liptai&Price_2019}, is conceptually straightforward and of significant interest. We leave this development for future works.

\section*{Acknowledgment}
This work was supported by JSPS KAKENHI Grant Number 24KJ1259, 24KJ1302, and 25H00394.
 The numerical calculations were performed on the XC50 system at the Center for Computational Astrophysics (CfCA) of the National Astronomical Observatory of Japan.

\clearpage

\bibliography{SRGSPH}{}
\bibliographystyle{aasjournal}

\end{document}